\documentclass[12pt]{iopart}
\usepackage{graphicx}
\usepackage{iopams}
\expandafter\let\csname equation*\endcsname\relax
\expandafter\let\csname endequation*\endcsname\relax
\usepackage{amsmath}
\usepackage{amssymb}
\usepackage{csquotes}
\MakeOuterQuote{"}
\usepackage{bm}
\usepackage{pgfplots}
\usepackage{tikz}
\usepackage{hyperref}
\usepackage[capitalise]{cleveref}  
\usepackage[acronym]{glossaries-extra}

\crefname{figure}{Fig.}{Figs.}
\crefname{equation}{Eq.}{Eqs.}
\crefname{section}{Sec.}{Secs.}
\glsdisablehyper
\setabbreviationstyle[acronym]{long-short}
\newacronym{re}{RE}{runaway electron}
\newacronym{mhd}{MHD}{magnetohydrodynamics}
\newacronym{cfl}{CFL}{Courant–Friedrichs–Lewy}
\newacronym{vde}{VDE}{vertical displacement event}
\newacronym{pde}{PDE}{partial differential equation}
\newacronym{ode}{ODE}{ordinary differential equation}
\newacronym{rk4}{RK4}{4th order Runge-Kutta}
\newacronym{hfs}{HFS}{high field side}
\newacronym{mgi}{MGI}{massive gas injection}
\newacronym{ma}{MA}{mega-ampera}
\newacronym{gmres}{GMRES}{generalized mininal residual method}
\newacronym{lcfs}{LCFS}{last closed flux surface}

\begin{document}

\title{Self-consistent simulation of resistive kink instabilities with runaway electrons}

\author{Chang Liu}
\address{Princeton Plasma Physics Laboratory, Princeton, NJ, United States of America}
\ead{cliu@pppl.gov}
\author{Chen Zhao}
\address{Princeton Plasma Physics Laboratory, Princeton, NJ, United States of America}
\author{Stephen C. Jardin}
\address{Princeton Plasma Physics Laboratory, Princeton, NJ, United States of America}
\author{Nathaniel Ferraro}
\address{Princeton Plasma Physics Laboratory, Princeton, NJ, United States of America}
\author{Carlos Paz-Soldan}
\address{Columbia University, New York, NY, United States of America}
\author{Yueqiang Liu}
\address{General Atomics, San Diego, CA, United States of America}
\author{Brendan C. Lyons}
\address{General Atomics, San Diego, CA, United States of America}

\begin{abstract}
	
A new fluid model for runaway electron simulation based on fluid description is introduced and implemented in the magnetohydrodynamics code M3D-C1, which includes self-consistent interactions between plasma and runaway electrons. The model utilizes the method of characteristics to solve the continuity equation for the runaway electron density with large convection speed, and uses a modified Boris algorithm for pseudo particle pushing. The model was employed to simulate magnetohydrodynamics instabilities happening in a runaway electron final loss event in the DIII-D tokamak. Nonlinear simulation reveals that a large fraction of runaway electrons get lost to the wall when kink instabilities are excited and form stochastic field lines in the outer region of the plasma. Plasma current converts from runaway electron current to Ohmic current. Given the good agreement with experiment, the simulation model provides a reliable tool to study macroscopic plasma instabilities in existence of runaway electron current, and can be used to support future studies of runaway electron mitigation strategies in ITER.

\end{abstract}

\submitto{\PPCF}

\maketitle

\section{Introduction}

Severe damage can be caused by high-energy \glspl{re} generated in tokamak disruption events, which is one of the major threats to the safe operation of ITER \cite{lehnen_disruptions_2015}. It is predicted that large populations of \glsps{re} can be generated during the current quench phase through knock-on collisions and the resulting \gls{re} avalanche \cite{boozer_theory_2015,martin-solis_formation_2017}. The current associated with \glsps{re} can be several \glspl{ma}. It can alter the macroscopic \gls{mhd} stability conditions and thus plays an important role in the disruption process. Several present-day tokamaks, including DIII-D \cite{paz-soldan_recent_2019}, JET \cite{reux_runaway_2015}, ASDEX \cite{kwon_study_1988}, and J-TEXT \cite{zeng_runaway_2017} have been used to test \gls{re} avoidance and mitigation strategies in order to control this issue in ITER and future fusion reactors.

In recent experiments on DIII-D \cite{paz-soldan_kink_2019} and JET \cite{reux_demonstration_2021} with large \gls{re} current generation, significant \gls{mhd} instabilities are observed in the \gls{re} current plateau phase after the initial disruption, which leads to the loss of most \glsps{re} within tens of microseconds . These experiments indicate the importance of \gls{mhd} instabilities in a successful \gls{re} mitigation. In the experiments, high-$Z$ impurities are expelled via deuterium injection, which also lowers the plasma density \cite{hollmann_study_2020}. The interaction between \gls{mhd} instabilities and \gls{re} current has been studied before theoretically with both analytical theory and numerical simulations\cite{helander_resistive_2007,cai_influence_2015,bandaru_simulating_2019,zhao_simulation_2020,liu_structure_2020}. In the simulation, a fluid description of \glsps{re} is used to simplify the calculation, in which \gls{re} current is calculated from \gls{re} density, and the feedback of \gls{re} current to \gls{mhd} is included in the generalized Ohm’s law of \gls{mhd} equations. However, since \glsps{re} have a convection speed much larger than the Alfvén velocity, solving the continuity equation for the \gls{re} density is challenging, and can lead to numerical instabilities. To overcome this issue, a reduced value of the convection speed is often used, and artificial diffusion of \gls{re} density is introduced to damp high-$k$ modes.

In this paper, we present a new model for coupling \gls{re} in \gls{mhd} simulations, and discuss its implementation in M3D-C1. This model is based on the fluid description of \glspl{re} \cite{zhao_simulation_2020}. The convection part of the \gls{re} continuity equation is treated using the method of characteristics, which can help avoid numerical instabilities associated with the \gls{cfl} condition. The solution of the characteristic equation can be obtained by following pseudo particles along  characteristic lines, which is similar to particle-pushing in a particle-in-cell (PIC) simulation and can be easily accelerated using parallel computing and GPUs. In addition, we utilized a modified Boris algorithm to advance pseudo particles, which can help limit the accumulation of numerical error. This new simulation model enables us to efficiently calculate the \gls{re} continuity equation with a large value of the convection speed, comparable to the speed of light.

Using this model we simulate a resistive kink instability happening in the \gls{re} plateau. The simulation is based on a DIII-D shot 177040, in which a large \gls{re} current is generated from the avalanche after the initial disruption, and finally leads to a “second disruption” when the edge safety factor ($q_a$) approaches 2, and causes the sudden loss of all \glsps{re} \cite{paz-soldan_kink_2019}. The linear simulation shows the dominance of the (2,1) kink mode near the edge, in agreement with previous work using MARS-F \cite{liu_mars-f_2019}. In the nonlinear simulation, it is found that as the (2,1) resistive kink mode grows exponentially, more than $95\%$ of the \glspl{re} can get lost to the wall due to the breaking of flux surfaces and formation of stochastic field lines. The plasma current is converted from \gls{re} current to Ohmic current, and its Ohmic heating effect results in an increase of plasma temperature, which is self-consistently modeled in the simulation. After the initial strike of \gls{mhd} instability, the outer region of the plasma remains stochastic, and the current density near the magnetic axis increases which drives a (1,1) kink mode and flattens the residual \gls{re} density in the core region.

This paper is organized as follows. In \cref{sec:fluid-model} we introduce the fluid model of \glspl{re} that was implemented in M3D-C1. In \cref{sec:characteristic-method} we discuss the motivation and implementation of the method of characteristics for solving the \gls{re} convection equation. In \cref{sec:boris-algorithm} we discuss the modified Boris algorithm used in pseudo particle pushing, and illustrate its conservation property. In \cref{sec:numerical-simulation} we discuss the simulation of a resistive kink instability with \glspl{re} happening in a DIII-D equilibrium, including the linear growth rate and real frequency, and nonlinear saturation and \gls{re} loss. In \cref{sec:summary} we give the conclusions and discussion.

\section{Fluid model of \gls{re} in \gls{mhd}}
\label{sec:fluid-model}

M3D-C1 is an initial value code that solves the 3D \gls{mhd} equations in tokamak  geometry \cite{ferraro_calculations_2009}. The code utilizes high-order $C^1$ continuous finite elements on a 3D mesh which is unstructured in $(R, Z)$ but extruded in the toroidal angle. It has the options to evolve the equations using fully-implicit or semi-implicit methods \cite{jardin_high-order_2007}. The sparse matrices generated from the Galerkin method are typically solved using the \gls{gmres} method with a block-Jacobi preconditioner. In addition to the plasma region, the code can also contain a resistive wall region and a vacuum region, and can represent halo currents shared by the plasma and wall regions \cite{ferraro_multi-region_2016}. In combination with a kinetic code, M3D-C1 can also be used to study  effects such as neoclassical tearing modes \cite{lyons_steady-state_2015} and excitation of Alfvén modes driven by energetic particles.

M3D-C1 has been used to study several phenomena related to tokamak disruptions, such as \gls{vde} \cite{clauser_vertical_2019}, generation of halo current \cite{pfefferle_modelling_2018}, and thermal quench due to impurities injection \cite{lyons_axisymmetric_2019}. Given the importance of \glspl{re} in disruption studies, we have implemented a fluid model of \glspl{re} in M3D-C1, by adding a new equation describing the evolution of \gls{re} density \cite{zhao_simulation_2020}. The \gls{re} current is calculated using \gls{re} density based on a simplified model of \gls{re} momentum distribution. The whole set of \gls{mhd} equations solved by M3D-C1 can be written as,
\begin{equation}
	\label{eq:momentum}
	n m\left[ \frac{\partial\mathbf{V}}{\partial t}+\left(\mathbf{V}\cdot\nabla\right)\mathbf{V}\right]=e n_{RE}\mathbf{E}+\left(\mathbf{J}-\mathbf{J}_{RE}\right)\times\mathbf{B}-\nabla p,
\end{equation}
\begin{equation}
	\label{eq:density}
	\frac{\partial n}{\partial t}+\nabla\cdot\left(n \mathbf{V}\right)=0,
\end{equation}
\begin{equation}
	\label{eq:re-convection}
	\frac{\partial n_{RE}}{\partial t}+\nabla\cdot\left[n_{RE}\left(c_{RE}\mathbf{b}+\frac{\mathbf{E}\times\mathbf{B}}{B^2}\right)\right]=S_{RE},
\end{equation}
\begin{equation}
	\label{eq:jre}
	\mathbf{J}_{RE}=-e n_{RE} \left(c \mathbf{b} +\frac{\mathbf{E}\times\mathbf{B}}{B^2}\right),
\end{equation}
\begin{equation}
	\label{eq:ohmslaw}
	\mathbf{E}=-\mathbf{V}\times\mathbf{B}+\eta\left(\mathbf{J}-\mathbf{J}_{RE}\right),
\end{equation}
\begin{equation}
	\label{eq:faraday}
	\frac{\partial \mathbf{B}}{\partial t}=-\nabla\times\mathbf{E},\quad \nabla\times\mathbf{B}=\mathbf{J},
\end{equation}
\begin{equation}
	\label{eq:pressure}
	p=n(T_e+T_i),
\end{equation}
\begin{equation}
	\label{eq:temperature}
	\frac{n}{\left(\gamma-1\right)}\left[\frac{\partial \left(T_e+T_i\right)}{\partial t}+\nabla\cdot\left(\left(T_e+T_i\right)\mathbf{V}\right)\right]=-n\left(T_e+T_i\right)\nabla\cdot\mathbf{V}-\nabla\cdot\mathbf{q}+\eta (\mathbf{J}-\mathbf{J}_{RE})^2,
\end{equation}
\begin{equation}
	\label{eq:heatflux}
	\mathbf{q}=n\left(\kappa_\perp\bm{\nabla} +\mathbf{b}\kappa_\parallel\nabla_\parallel\right)\cdot\left(T_e+T_i\right).
\end{equation}
In this set, \cref{eq:momentum} is the \gls{mhd} momentum equation. $n$ is the plasma number density, $m$ is the ion mass, $\mathbf{V}$ is \gls{mhd} velocity, $\mathbf{E}$ is the electric field, $e$ is the elementary charge, $\mathbf{J}$ is the total current, $\mathbf{B}$ is the magnetic field, and $p$ is the plasma pressure. The first term on the right-hand-side is due to the positive charge of plasma if excluding \glsps{re}. \cref{eq:density} is the continuity equation of plasma density. \cref{eq:re-convection} is the continuity  equation for \gls{re} density $n_{RE}$, where $\mathbf{b}=\mathbf{B}/B$, $c_{RE}$ is the \gls{re} convection velocity, and $S_{RE}$ is a source term representing \gls{re} generation. \gls{re} current $\mathbf{J}_{RE}$ is calculated from $n_{RE}$ in \cref{eq:jre}, where $c$ is the speed of light. \cref{eq:ohmslaw} is the generalized Ohm’s law where $\eta$ is the plasma resistivity. \cref{eq:faraday} is the Faraday’s law, which is used to advance the vector potential $\mathbf{A}$ so that $\mathbf{B}=\nabla\times\mathbf{A}$ remains divergence free. The pressure $p$ is calculated from the electron temperature $T_e$ and ion temperature $T_i$. Here we use a unified temperature for both assuming $T_e=T_i$. \cref{eq:temperature} is the temperature equation where $\gamma$ is the ratio of specific heats and $\mathbf{q}$ is the heat flux. The last term characterizes Ohmic heating. $\mathbf{q}$ can be calculated with \cref{eq:heatflux}, where $\kappa_\parallel$ and $\kappa_\perp$ are parallel and perpendicular heat conduction coefficients.

\cref{eq:re-convection} describes streaming of \glspl{re} along magnetic field lines with parallel velocity $c_{RE}$ and  $\mathbf{E}\times\mathbf{B}$ drift included. Given that most of the \glspl{re} are relativistic particles with a very small pitch angle, the value of $c_{RE}$ should be close to the speed of light $c$, but is usually set to be a smaller value for numerical efficacy. Note that here the gradient and curvature drifts of \glspl{re} are not included, assuming that the average energy of \glspl{re} is small thus these drift motions are subdominant compared to parallel streaming.

The current formed by \glspl{re} can be represented as \cref{eq:jre}. When substituting this $\mathbf{J}_{RE}$ into \cref{eq:momentum}, it is found that the component perpendicular to $\mathbf{B}$ can cancel the $-\mathbf{J}_{RE}\times\mathbf{B}$ term, and only the parallel component is left. This component is not included in the current implementation, given that runaway density is much smaller compared to that of thermal electrons and $\mathbf{E}$ parallel to $\mathbf{B}$ is smaller than perpendicular part. Since the collisional friction of \glspl{re} can be ignored compared to thermal electrons, the resistive term in the generalized Ohm’s law is reduced as in \cref{eq:ohmslaw}. 

In the split time advance scheme in M3D-C1, the \gls{mhd} equations are calculated following the orders in Eq.~(\ref{eq:momentum}-\ref{eq:heatflux}) for every \gls{mhd} timestep. For each equation the time-integration quantity is solved using $\theta$-implicit method. The momentum equation is advanced using the semi-implicit method, by including a parabolic term that was derived from second order time derivatives, to ensure numerical stability \cite{jardin_high-order_2007}. The plasma and \gls{re} densities are then calculated using the magnetic field and electric field from the previous timestep. The magnetic field equation and the pressure equation are advanced at the last step, using the velocities and \gls{re} density obtained from both the current and previous timesteps.

\section{Method of characteristics for solving \gls{re} convection}
\label{sec:characteristic-method}

As discussed in the \cref{sec:fluid-model}, the continuity equation of \gls{re} density (\cref{eq:re-convection}) needs to be solved for the fluid \gls{re} model. In the previous implementation used for linear \gls{mhd} simulation in a 2D mesh including \gls{re} current \cite{zhao_simulation_2020}, this equation was solved using the same numerical method as other \gls{mhd} equations. First a sparse matrix is constructed by way of the $\theta$-implicit method and the Galerkin method. Then this matrix is solved using a direct solver.

However, it becomes more challenging to solve the \gls{re} continuity equation in a 3D mesh for nonlinear simulation with $c_{RE}\gg v_A$, where $v_A$ is the Alfvén velocity. In 3D, a direct solver is no longer feasible, and the \gls{gmres} iterative method is used with a block-Jacobi preconditioner.  This preconditioner is not optimal for \cref{eq:re-convection} because the second term is dominated by the toroidal derivative of the \gls{re} density ($\partial_\varphi n_{RE}$, where $\varphi$ is the toroidal angle) as $\mathbf{b}$ is mainly along the toroidal direction. In the linear simulation where the toroidal derivative can be represented using a Fourier mode number, this term is a diagonal term in the matrix. In nonlinear simulation using 3D mesh, this term becomes an off-diagonal term with block-Jacobi method and can dominate the matrix when $c_{RE}\gg v_A$ if the timestep is on the order of the Alfvén time. To overcome the singularity and make the matrix solver converge, one can use a smaller timestep and introduce subcycles when solving this continuity equation, which will increase the computation time. Additionally one can use a smaller value of $c_{RE}$ to reduce the singularity. This method was used in  \cite{cai_influence_2015,bandaru_simulating_2019} and in our previous linear simulation \cite{zhao_simulation_2020}, where we showed that the linear growth rate of \gls{mhd} modes is not sensitive to the value of $c_{RE}$. However, it is found in our nonlinear simulation that one needs a large enough value of $c_{RE}$ to get a converged result of mode saturation amplitude, as discussed in \cref{sec:nonlinear-simulation}.

Another issue in our previous implementation is numerical instability, which is linked to the singularity problem and can become serious as $c_{RE}$ becomes large. This is because \cref{eq:re-convection} is a pure convection equation and high $k$ modes caused by numerical error will not diffuse. This issue can be overcome by introducing an artificial diffusion term into \cref{eq:re-convection}, which was used in in previous work. However, when $c_{RE}$ is large, this diffusion term also needs to be large to suppress the numerical instability, and this large diffusion term can possibly reduce the \gls{mhd} mode growth or even suppress unstable modes.

In view of the above issues, we developed a new method to deal with \gls{re} convection, by converting the convection part of \cref{eq:re-convection} from a \gls{pde} into an \gls{ode} and solve it using the method of characteristics. Using some properties of magnetic fields, the convection part of \cref{eq:re-convection},
\begin{equation}
	\label{eq:re-convection2}
	\frac{\partial n_{RE}}{\partial t}+\nabla\cdot\left[n_{RE}\left(c_{RE}\mathbf{b}+\frac{\mathbf{E}\times\mathbf{B}}{B^2}\right)\right]=0,
\end{equation}
can be rewritten in a divergence-free form,
\begin{equation}
	\label{eq:re-convection3}
	\frac{\partial}{\partial t}\left(\frac{n_{RE}}{B}\right)+\left(c_{RE}\mathbf{b}+\frac{\mathbf{E}\times\mathbf{B}}{B^2}\right)\cdot\nabla\left(\frac{n_{RE}}{B}\right)=\frac{n_{RE}}{B^2}\left[\mathbf{E}\cdot\left(\nabla\times\mathbf{b}\right)\right].
\end{equation}
The details of this conversion can be found in \ref{sec:divergence-free}. The last term characterizes the change of $n_{RE}/B$ due to the horizontal displacement motion and resistive diffusion, both of which happen on the resistive timescale that is much longer than the time for \gls{re} convection along field lines. This term can be treated as an extra source term, and the rest of the equation can be easily solved by following the characteristic line,
\begin{equation}
	\label{eq:characteristic-line}
\frac{d\mathbf{x}(t)}{dt}=c_{RE}\mathbf{b}+\frac{\mathbf{E}\times\mathbf{B}}{B^2},
\end{equation}
which can help convert the \gls{pde} into an \gls{ode},
\begin{equation}
	\frac{d}{dt}\left(\frac{n_{RE}}{B}\right)\left[\mathbf{x}(t),t\right]=\frac{\partial}{\partial t}\left(\frac{n_{RE}}{B}\right)\left[\mathbf{x}(t),t\right]+\frac{d\mathbf{x}(t)}{dt}\cdot\nabla\left(\frac{n_{RE}}{B}\right)\left[\mathbf{x}(t),t\right]=0,
\end{equation}
whose solution is just $\left(n_{RE}/B\right)\left[\mathbf{x}(t_0),t_0\right]=\left(n_{RE}/B\right)\left[\mathbf{x}(t=0),0\right]$. Therefore, to solve the convection equation, one needs to integrate along the characteristic line (\cref{eq:characteristic-line}) backwards in time to obtain the value of $\left(n_{RE}/B\right)$ at $\mathbf{x}(t=0)$, and then do a pull-back transformation to obtain the new value of $\left(n_{RE}/B\right)$ from it.

When applying this method, we are using the new field quantity  $\left(n_{RE}/B\right)$ instead of $n_{RE}$ to represent \gls{re} density. In M3D-C1, all the field quantities are described as coefficients of finite element basis functions. These coefficients can be calculated by solving a linear algebra equation, if the value of the fields at each quadrature points is known,
\begin{equation}
	\label{eq:mass-matrix}
	\sum_i \left(\nu_i, C_i u_i\right)=\sum_i \left(\nu_i, \frac{n_{RE}}{B}\right)=\sum_i \sum_m \nu_i(\mathbf{x}=\mathbf{x}_m) \frac{n_{RE}}{B}(\mathbf{x}=\mathbf{x}_m)J(\mathbf{x}=\mathbf{x}_m),
\end{equation}
where $\nu_i$ is the $i$-th test function and $u_i$ is the basis function. $C_i$ is the coefficient corresponding to $u_i$ for the field $\left(n_{RE}/B\right)$. $x_m$ is the $m$-th quadrature point used for integration. $(\cdot,\cdot)$ is the inner product and $J$ is the Jacobian for spatial integration. The equation on the left-hand-side can be written as $\mathbf{M}\cdot\mathbf{C}$, where $\mathbf{M}$ is the mass matrix. Therefore, to obtain the values of $C_i$ of $\left(n_{RE}/B\right)$, one needs to calculate its value at each quadrature point at the end of the timestep. The calculation steps using the method of characteristics can be described as follows:
\begin{enumerate}
	\item Initialize “pseudo particles” at each quadrature point of every element in the whole 3D mesh.
	\item Push all the particle following \cref{eq:characteristic-line}, for a period of $\Delta t$ but backwards in time.
	\item Pull the value of $n_{RE}/B$ at the final position to the initial quadrature point.
	\item Solve the mass matrix equation \cref{eq:mass-matrix} to obtain the coefficients $C_i$ of $n_{RE}/B$ at the new timestep.
	\item Calculate the value of \gls{re} source term and add it to the field.
\end{enumerate}

This process is similar to advancing particles and calculating their density through mesh deposition in a PIC simulation. The difference is that in this method \gls{re} density is represented as a \gls{mhd} field rather than a moment of the distribution function. Therefore, there is no need to pump in new particles when \glsps{re} are lost, and the numerical noise due to the spikiness of the particle distribution is not present.

Since the \gls{re} continuity equation has been converted to an \gls{ode}, the choice of timestep for for pseudo particle pushing is not limited by the \gls{cfl} condition, and numerical instability can be avoided even with large $\Delta t$. This also means that the artificial diffusion term which was introduced in the \gls{re} continuity equation in previous work is not needed. Nevertheless, to ensure the accuracy of the \gls{re} orbit calculation with large $c_{RE}$, we choose a smaller timestep $\Delta t'$ for \cref{eq:characteristic-line} calculation than what is used for the \gls{mhd} calculation. This means that there are multiple subcycles of particle pushing within each \gls{mhd} timestep, during which the \gls{mhd} fields are assumed to be fixed.  Note that the pushing of pseudo particles are independent of each other, thus can be easily parallelized and accelerated using GPUs.

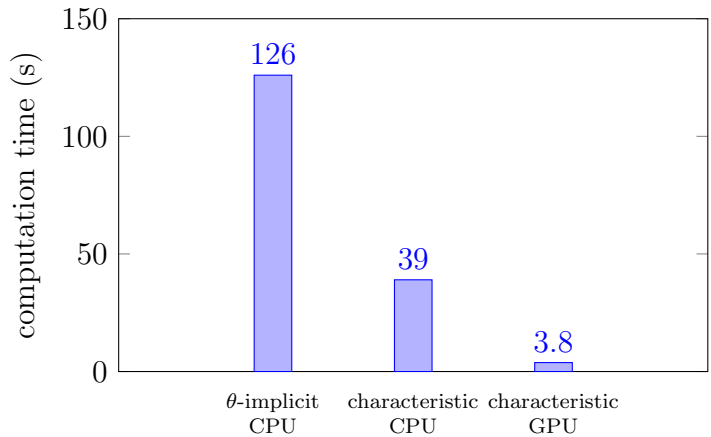
\begin{figure}[h]
	\begin{center}
    \begin{tikzpicture}
		\begin{axis}[
			ybar,
			enlarge x limits=0.55,
			bar width=.5cm,
			width=.6\textwidth,
			height=.4\textwidth,
			symbolic x coords={jacobi,cpu,gpu},
			xtick=data,
			xtick style={draw=none},
			xticklabel style={align=center,font=\scriptsize},
			xticklabels={$\theta$-implicit\\ CPU, characteristic\\ CPU,characteristic\\ GPU},
			nodes near coords,
			nodes near coords align={vertical},
			ymin=0,ymax=150,
			ylabel={computation time (s)},
		    ]
			\addplot coordinates {(jacobi,126) (cpu,39) (gpu,3.8)};
		\end{axis}
	\end{tikzpicture}
    \end{center}
    \caption{\label{fig:computation-time}Computation time for calculating \gls{re} convection equation for 1290800 pseudo particles and $\Delta t=0.65\mu$s with $c_{RE}/v_A=8$ using different methods and processors.}
\end{figure}

To illustrate the performance of the method of characteristics, we did a benchmark for calculating \cref{eq:re-convection2} using both the $\theta$-implicit method using block-Jacobi preconditioner and the new method of characteristics on CPUs and GPUs, for $\Delta t=0.65\mu$s). This benchmark is conducted using the Summit cluster. The CPU computation is done using 32 IBM POWER9 CPUs and the GPU computation is done using 96 NVIDIA Volta V100s GPUs. In the benchmark we employ a mesh with 16 toroidal planes and 6454 elements per plane. The \gls{re} convection velocity is set as $c_{RE}=8 v_A$, which is close to the value of speed of light. The total number of pseudo particles is 1290800, which is equal to the quadrature points of all elements (125 per element). Even with such large number of pseudo particles, the method of characteristics still outperforms the classical method on computation time, as shown in \cref{fig:computation-time}. The reason is that, to make the $\theta$-implicit method work with such a large value of $c_{RE}$, we need to use a large number of subcycles to make the \gls{gmres} solver converge, which slows down the solving of the continuity equation significantly. For the method of characteristics, we apply $\Delta t'=\tau_A/32$ for pseudo particle pushing, but the calculation is still faster given the simplicity of particle pushing operation and the acceleration brought by parallelization computing, especially on GPUs. This example shows that for simulations with large $c_{RE}$, the method of characteristics has a significant advantage over the classical method.

\section{Modified Boris algorithm for pseudo particle trajectory calculation}
\label{sec:boris-algorithm}

When using the method of characteristics to solve the \gls{re} continuity equation, one needs to integrate the trajectories of pseudo particles to update the value of the \gls{re} density. It is known that classical integration methods for solving \glspl{ode}, such as the explicit Runge-Kutta method, do not conserve physical quantities and suffer from accumulation of numerical error for long time simulations. This problem is more serious for the calculation of trajectories of \glsps{re} with large convection velocity. Inaccuracies of numerical integration can lead to the deviation of a pseudo particle trajectory from its original flux surface even without perturbations, and can break the conservation of \gls{re} density.

This error accumulation problem can be mitigated by using sympletic or structure-preserving integrators. Recently, a new algorithm using the idea of slow manifold for calculating the orbits of charged particles in magnetic fields has been developed \cite{xiao_slow_2020}. In this algorithm, the fast gyro motion of magnetized particles is ignored and only the slow manifold of motion is calculated, which is similar to the guiding center model. Structure-preserving algorithms such as the Boris algorithm can then be used to evolve this slow manifold with a timestep not limited by the gyro period. In order to take into account the effect gyro motion of the slow manifold, the mirror force ($-\mu\nabla B$) is introduced as an effective electric force on the particles. This method was shown to have good long time conservation properties, and was recently implemented in a MHD-kinetic hybrid code M3D-C1-K for particle pushing.

Inspired by this algorithm, we developed a modified Boris algorithm to advance pseudo particles following \cref{eq:characteristic-line}, which is structure-preserving and can be used for long time simulations. We start from the classical Boris algorithm, which can be described by the following equations,
\begin{equation}
	\label{eq:x-advance}
	\mathbf{x}_{l+1}=\mathbf{x}_{l}+\mathbf{v}_{l+1/2} \Delta t',
\end{equation}
\begin{equation}
	\label{eq:v-advance}
	\mathbf{v}'_{l+1/2}=\mathbf{v}_{l+1/2}-\frac{\mathbf{E}_{l+1}\times \mathbf{B}_{l+1}}{B^2},
\end{equation}
\begin{equation}
	\label{eq:v-rotation}
	\mathbf{v}^\dagger_{l+3/2}=\mathbf{v}'_{l+1/2}+\frac{q}{m}\left(\frac{\mathbf{v}'_{l+3/2}+\mathbf{v}'_{l+1/2}}{2}\times \mathbf{B}_{l+1}\right)\Delta t',
\end{equation}
\begin{equation}
	\label{eq:v-advance2}
	\mathbf{v}_{l+3/2}=\mathbf{v}^\dagger_{l+3/2}+\frac{\mathbf{E}_{l+1}\times \mathbf{B}_{l+1}}{B^2},
\end{equation}
where we subtract the $\mathbf{E}\times\mathbf{B}$ drift in \cref{eq:v-advance} before doing the velocity rotation with respect to $\mathbf{B}$ in \cref{eq:v-rotation}, and add it back after the rotation in \cref{eq:v-advance2}. Note that \cref{eq:characteristic-line} can be regarded as the equation of motion of massless particle in electromagnetic fields, where particles are only affected by $\mathbf{E}\times\mathbf{B}$ drifts and not affected by gradient or curvature drifts which depend on particle mass. In terms of that, we can take the limit of $m\to 0$ to obtain a modified Boris algorithm for integration of \cref{eq:characteristic-line}. In this limit, in order to satisfy \cref{eq:v-rotation}, the term in the parentheses must then vanish. Therefore \cref{eq:v-rotation} should be replaced by
\begin{equation}
	\left(\mathbf{v}'_{l+3/2}+\mathbf{v}'_{l+1/2}\right)\times \mathbf{B}_{l+1}=0,\qquad \left\lvert\mathbf{v}'_{l+3/2}\right\rvert=\left\lvert\mathbf{v}'_{l+1/2}\right\rvert.
\end{equation}
The last equation is added because the magnetic field force does not change the kinetic energy of particles. We can see that in this modified Boris algorithm, the magnetic field is not used as a guidance for the next step motion like in Runge-Kutta method. Instead, the next step velocity depends on the its value at the previous timestep and the magnetic field works like a reflection mirror that only reverses the component of $\mathbf{v}$ which is normal to it.

As discussed in \cite{xiao_slow_2020}, the Boris algorithm can be used to push particles with a timestep larger than the gyroperiod, as long as it follows the  slow manifold of particle motion without gyromotion. When integrating \cref{eq:characteristic-line}, we are only interested in drift motion including $\mathbf{E}\times\mathbf{B}$ drifts, so it is valid to apply the modified Boris algorithm with $\Delta t'$ larger than the electron gyro period. To ensure that the trajectory stays in the slow manifold, we use the \gls{rk4} method to calculate the particle motion of the initial timestep from $\mathbf{x}_0$ to $\mathbf{x}_1$ following \cref{eq:characteristic-line}, and then obtain $\mathbf{v}_{1/2}$ according to \cref{eq:x-advance}. These quantities are used as initial values of the modified Boris algorithm.

\begin{figure}[h]
	\begin{center}
		\includegraphics[width=0.45\linewidth]{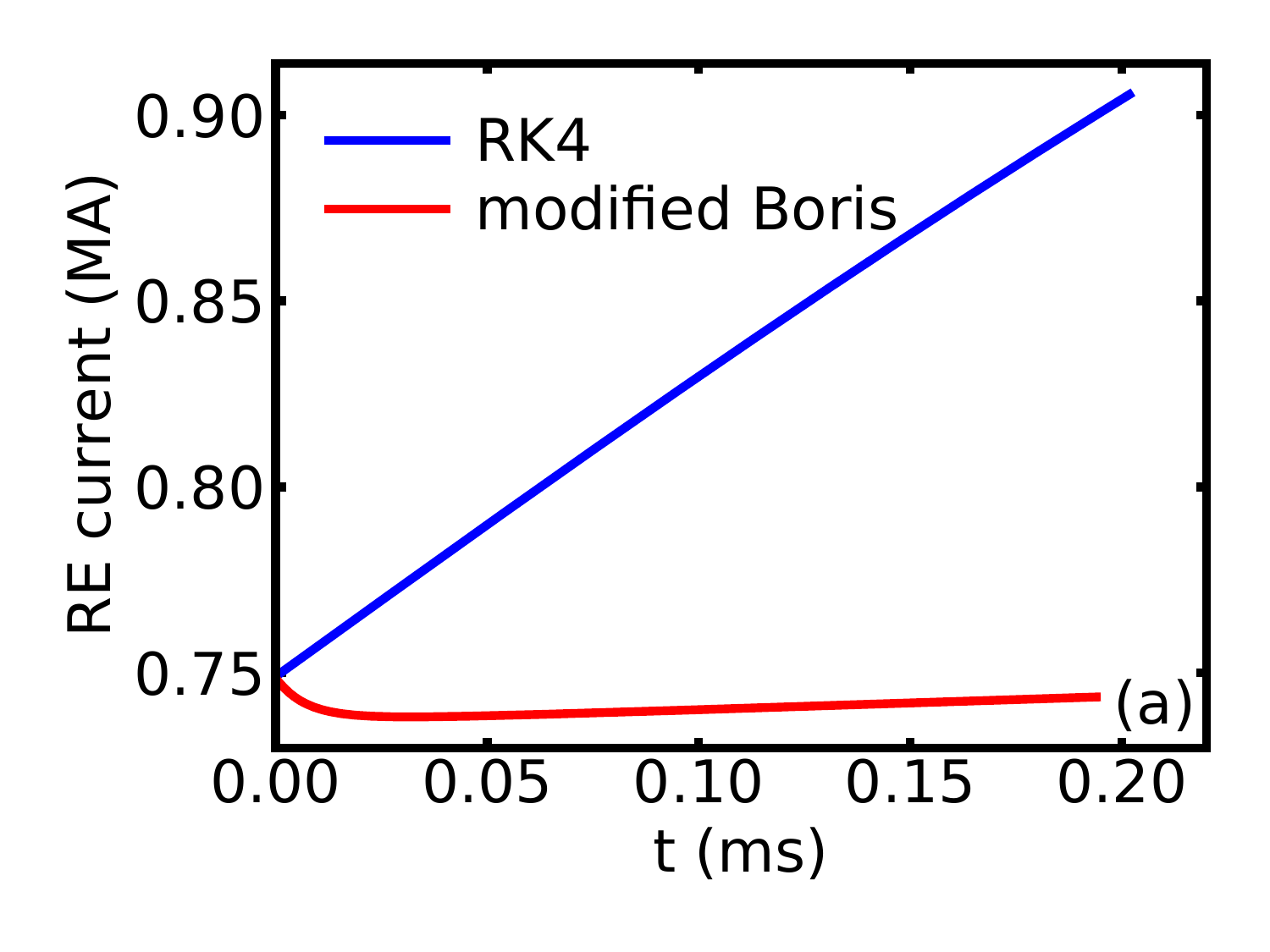}
		\includegraphics[width=0.48\linewidth]{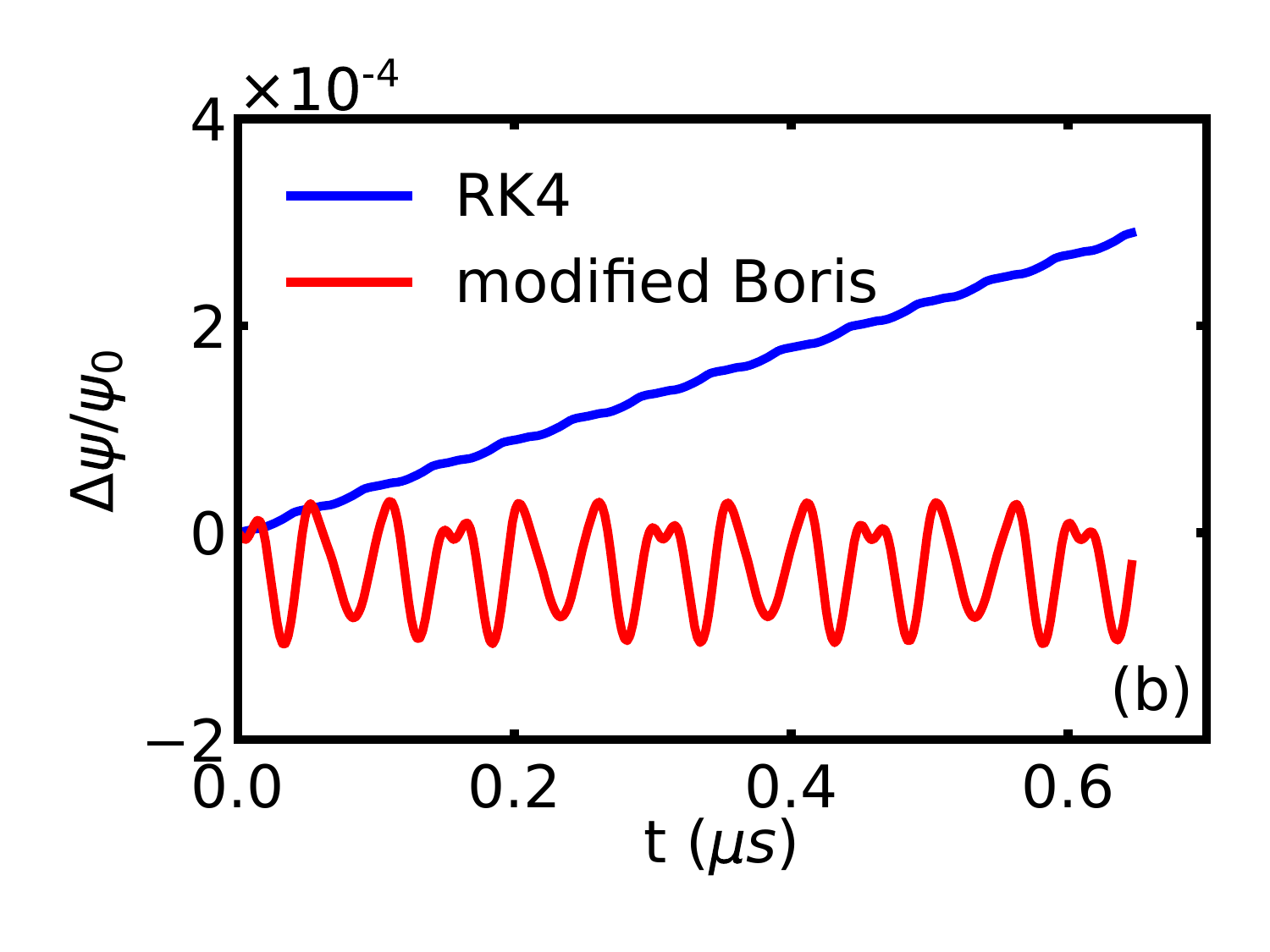}
	\end{center}
	\caption{\label{fig:boris-algorithm}(a) Time series of total \gls{re} current in a static field simulation with $c_{RE}/v_A=32$ using method of characteristics. The blue lines shows the result using \gls{rk4} for pseudo particle orbit calculation, and the red line shows the result using the modified Boris algorithm. (b) Conservation of magnetic flux for one pseudo particle using the two algorithms.}
\end{figure}

To show the conservation property of the modified Boris algorithm, we did a test simulation to compare the results of it  and \gls{rk4} for advancing \cref{eq:characteristic-line} with $c_{RE}=32v_A$. In this test, the \gls{re} current is advanced following \cref{eq:re-convection2} without electromagnetic perturbations or \gls{re} source. A plasma equilibrium from DIII-D shot 177040 is used, which is also used for the nonlinear simulation presented in \cref{sec:numerical-simulation}. Given that $\mathbf{J}=\mathbf{J}_{RE}$ and there is no resistivity associated with \gls{re} current, the total \gls{re} current should not change with time. \cref{fig:boris-algorithm} (a) shows the evolution of the total \gls{re} current using the two algorithms, where the change is due to the accumulation of numerical error. The error of the modified Boris algorithm is much smaller compared to that of \gls{rk4}. This result shows that the modified Boris algorithm provides better conservation properties for the \gls{re} density when calculating the \gls{re} continuity equation. Note that in this simulation the timestep for pseudo particle pushing of the modified Boris algorithm is 1/4 of the timestep of \gls{rk4}, as in \gls{rk4} the field evaluation needs to be done 4 times in one timestep of particle pushing. The total time for particle pushing using the two algorithms are almost the same.

To better understand the underlying reason for the difference between the results, we compare the conservation property of a single pseudo particle during the calculation. Given that the system is axisymmetric without any perturbation, the toroidal angular momentum $P_\varphi=mv_\varphi R+\psi_p$ should be conserved. Since pseudo particles are massless, there is no kinetic momentum associated with it and the conserved quantity is just $\psi_p$. \cref{fig:boris-algorithm} (b) shows the time evolution of $\psi_p$ for a single pseudo particle during the particle pushing, using two different algorithms. We can see that the total error of \gls{rk4} is larger for the period of integration. This error can lead to the start and end points of pseudo particle lying on different flux surfaces, resulting in an artificial diffusion and the breaking of \gls{re} density conservation. The numerical error of the modified Boris algorithm, though larger at each timestep, does not accumulate with time like in \gls{rk4}, which is consistent with the structure-preserving property of the Boris algorithm \cite{qin_why_2013}.

\section{Numerical simulation of resistive kink mode with \glspl{re} in DIII-D}
\label{sec:numerical-simulation}

In this section we show the simulation results using the \gls{re} module in M3D-C1 developed for \gls{mhd} instabilities happening in a high \gls{re} current equilibrium. This setup is based on DIII-D shot 177040, where a large \gls{re} current is driven by external loop voltage in the post-disruption phase. The \gls{re} current reaches about 1\gls{ma} and the safety factor of the \gls{lcfs} ($q_a$) drops to close to 2. Deuterium \gls{mgi} helps purge the impurities injected earlier to trigger the disruption, and reduces the plasma density. Note that \gls{mhd} instabilities can happen both during the \gls{re} current growing stage and the final loss stage when $q_a$ drops to 2. In the former stage, the instabilities happen intermittently, and the loss of \glsps{re} at the edge can be quickly compensated by the continuous generation near the core. In the simulation, we focus on the \gls{mhd} instabilities happening in the final loss event, where the majority of \glsps{re} are lost in a short time without regeneration.

\subsection{Plasma equilibrium}

\begin{figure}[h]
	\begin{center}
		\includegraphics[width=0.5\linewidth]{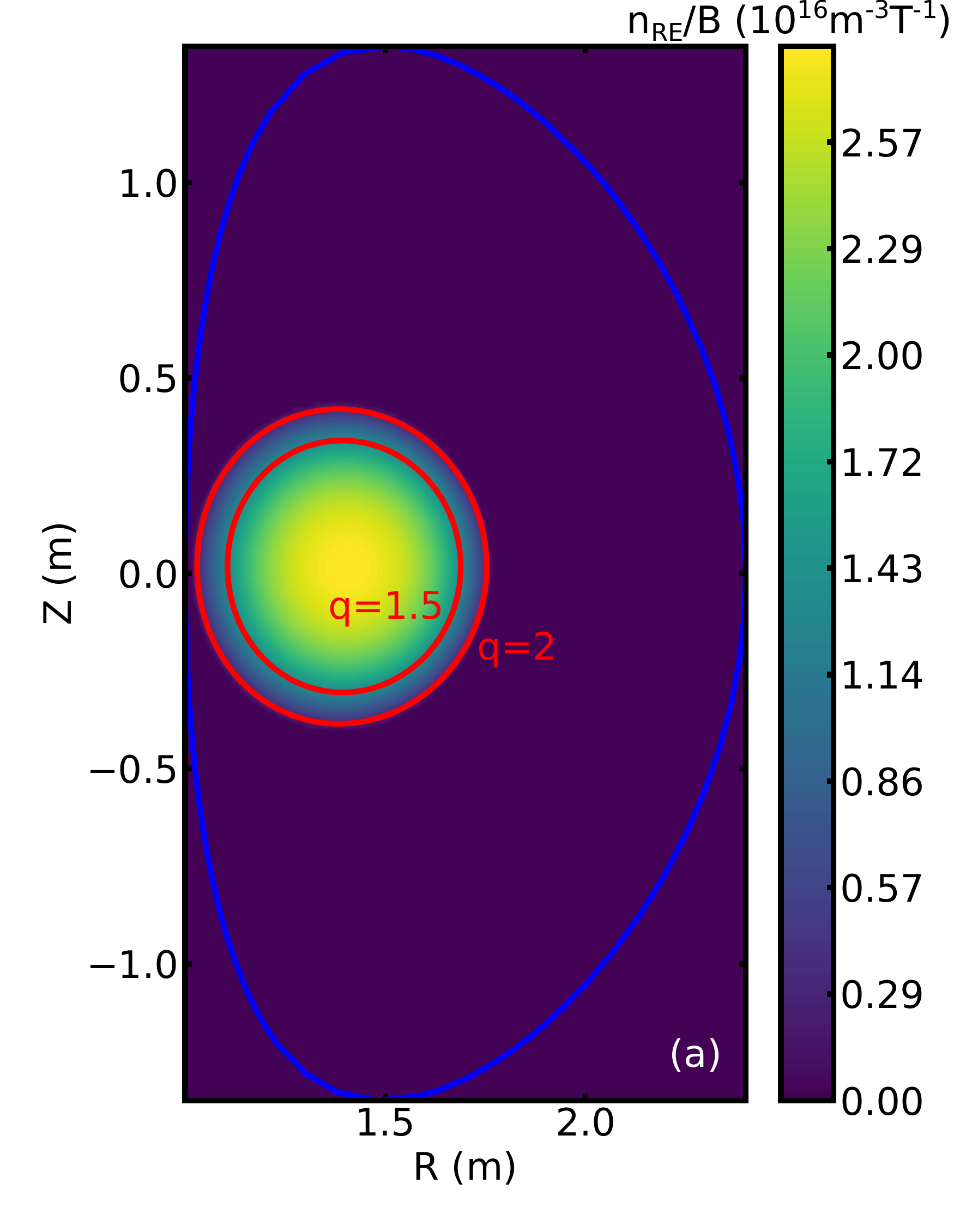}
		\includegraphics[width=0.43\linewidth]{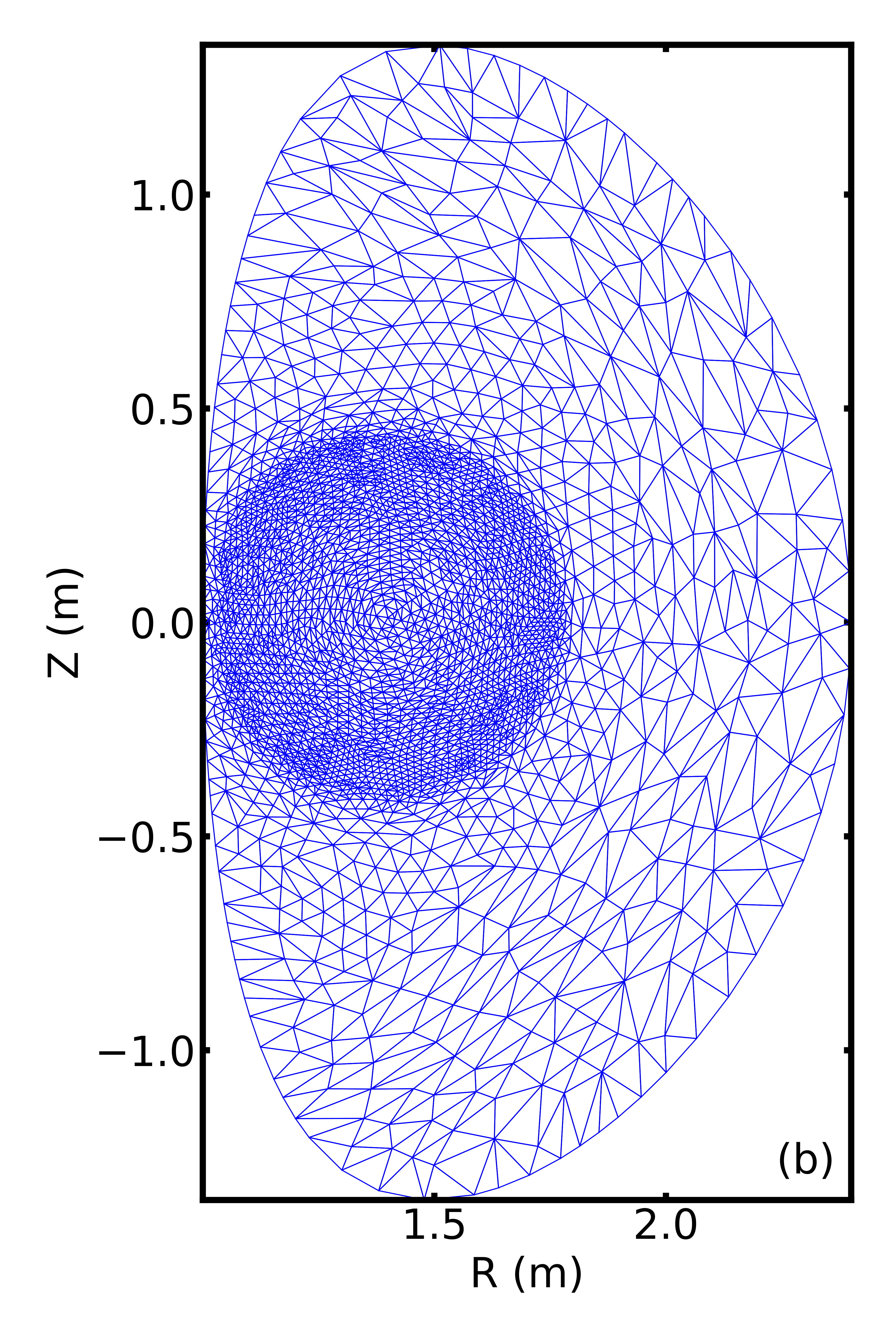}
	\end{center}
	\caption{\label{fig:RE-density-2d} (a) 2D profile of initial \gls{re} density. The red lines show the location of $q=1.5$ and $q=2$ flux surfaces. The blue lines shows the location of mesh boundary, which is also used as wall in the simulation for \gls{re} loss counting. (b) Finite element mesh used in M3D-C1 simulation.}
\end{figure}

Due to the \gls{re} shape control system available at DIII-D, the formed \gls{re} beam and the remaining closed flux surfaces are located near the \gls{hfs} in post-disruption plasma. The profiles of \gls{re} density, the flux contours, and the finite element mesh used for the simulation are shown in \cref{fig:RE-density-2d}. Note that the mesh density is high in the plasma region, especially near the $q=2$ flux surface in order to resolve the tearing layer structure.

\begin{figure}[h]
	\begin{center}
 		\includegraphics[width=0.5\linewidth]{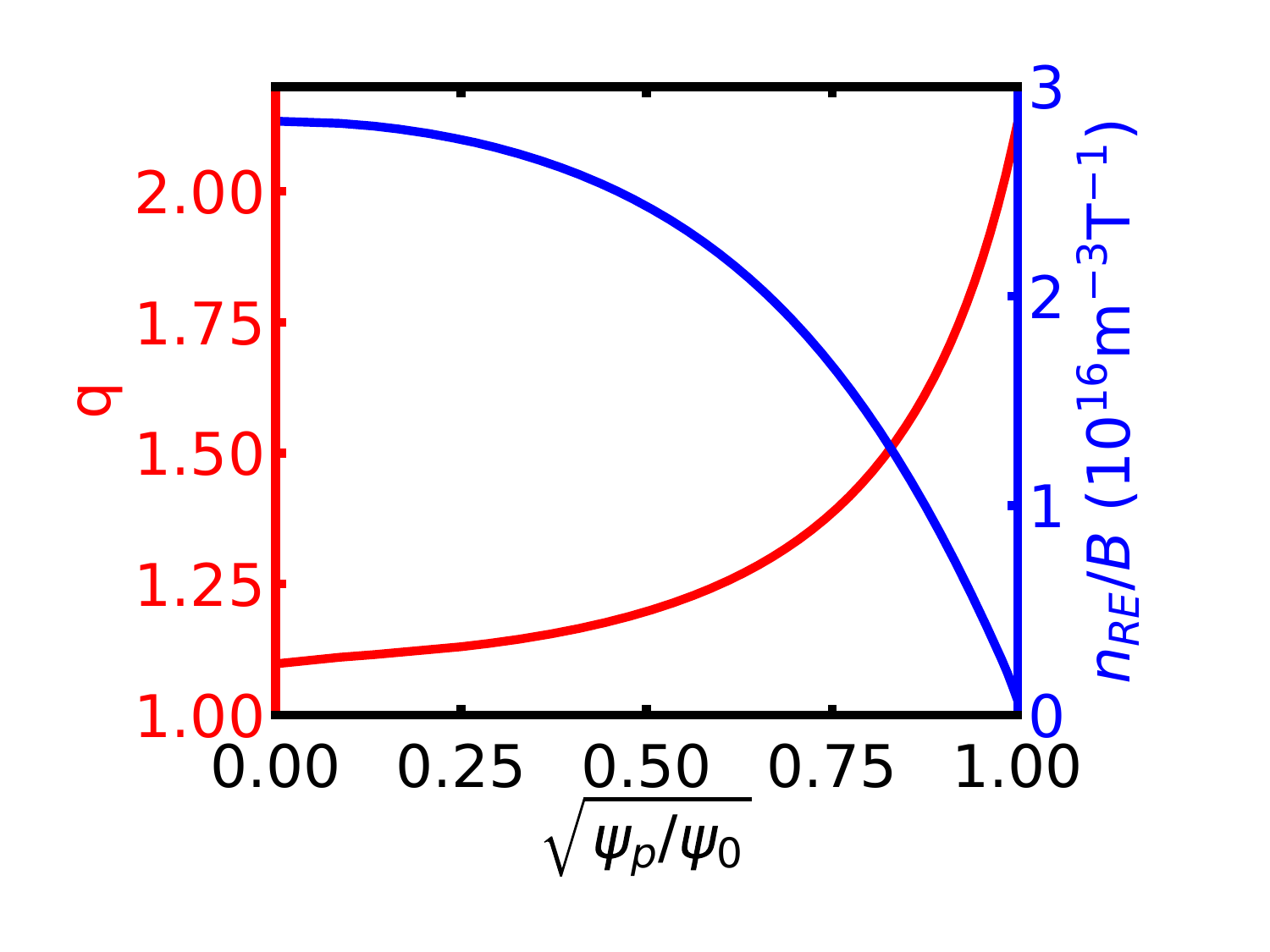}
	\end{center}
	\caption{\label{fig:q-profile}1D profiles of $q$ and $n_{RE}/B$, as functions of square root of normalized poloidal flux.}
\end{figure}

The initial plasma equilibrium satisfying the Grad-Shafranov equation was obtained using the equilibrium code EFIT with the experimental data at 1025ms, which is just before the final loss event happens. The equilibrium  pressure was set close to zero, thus the force balance equation can be simplified as $\mathbf{J}\times\mathbf{B}\approx 0$. This is to ensure that the equilibrium current $\mathbf{J}$ is almost parallel to $\mathbf{B}$, which is consistent with the fluid \gls{re} model by assuming all equilibrium current is carried by \glsps{re} ($\mathbf{J}=\mathbf{J}_{RE}$). The obtained new equilibrium is shown in \cref{fig:q-profile}, including the profile of safety factor ($q$) and $n_{RE}/B$. Note that the equilibrium RE current does not have an off-axis peak like JET equilibrium in \cite{bandaru_magnetohydrodynamic_2021}, and both $q$ and $n_{RE}$ are monotonic function of $\psi_p$. The toroidal field at the magnetic axis is $B_T=2$T. Plasma density is set at $n=1.5\times 10^{19}$m$^{-3}$. Ions are assumed to be all deuterium given that high-$Z$ impurities are expelled. Because of the low plasma density, the Alfvén velocity $v_A\approx 0.026 c$ which is larger than that in a normal tokamak discharge. $q_a$ of the initial equilibrium was about 2.1.

\subsection{Linear simulation of (2,1) resistive kink mode}
\label{sec:linear-simulation}

We first performed linear studies of the kink instability. This is done by running M3D-C1 in a 2D mesh with a spectral representation in the toroidal direction, assuming toroidal mode number $n=1$. The plasma resistivity is set to be uniform inside the \gls{lcfs}. Outside it the resistivity is set to be $10^3$ larger than the inside value in order in order to suppress current and simulate the  vacuum region \cite{ferraro_ideal_2010}. The value of $c_{RE}/v_A$ was set to be 8. Note that in a linear simulation, the \gls{re} density will be affected by the  perturbations of electromagnetic fields, which can give rise to $\delta n_{RE}$, but the \gls{re} characteristic line is calculated only using the equilibrium magnetic fields. Thus there is no \gls{re} loss in linear simulations.

\begin{figure}[h]
	\begin{center}
		\includegraphics[width=0.45\linewidth]{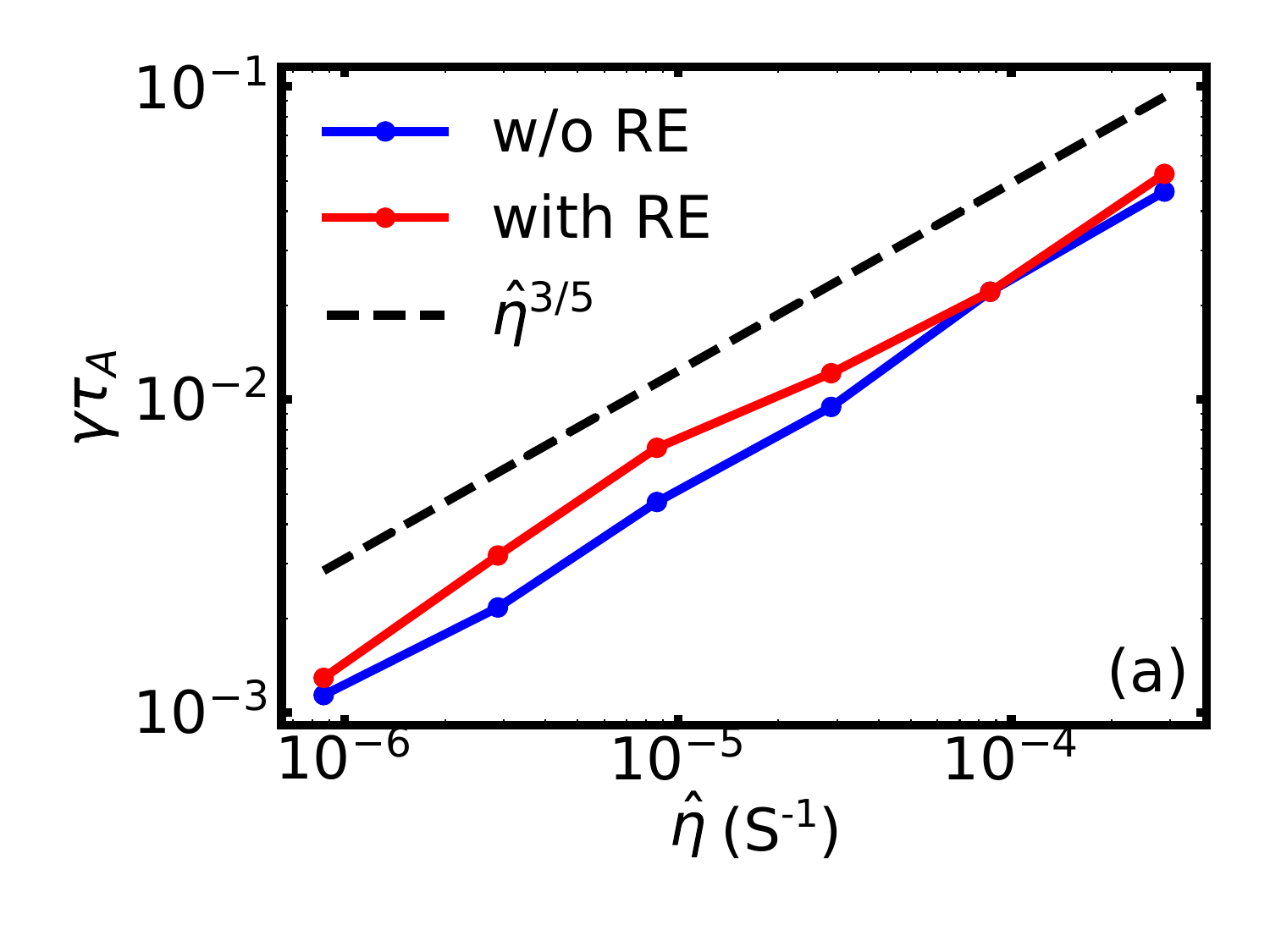}
		\includegraphics[width=0.45\linewidth]{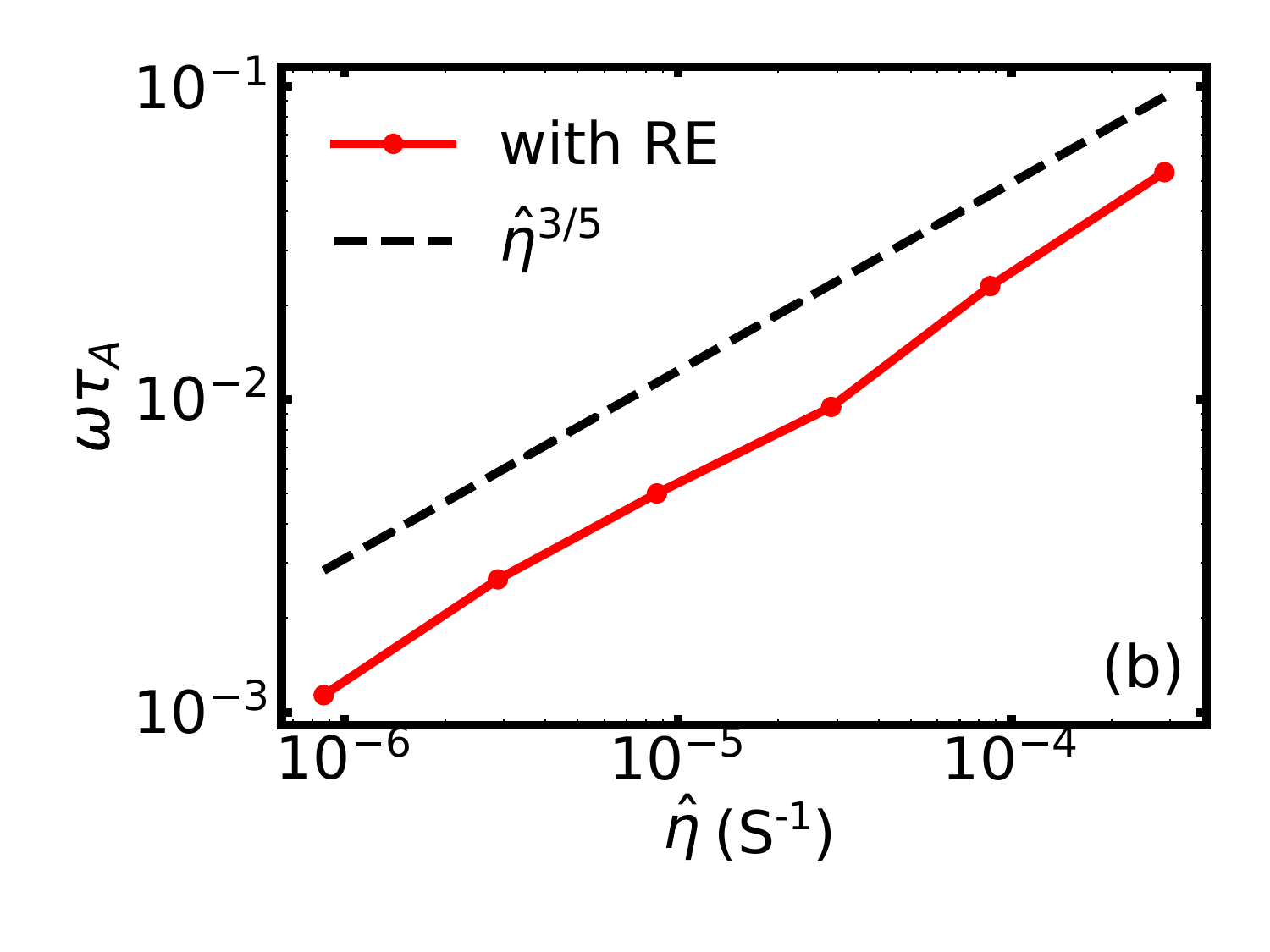}
	\end{center}
	\caption{\label{fig:linear-growthrate}Linear simulation results of (a) growth rate $\gamma$ and (b) real frequency $\omega$ of (2,1) resistive kink mode for different values of resistivity, for cases with $\mathbf{J}_{RE}=0$ (blue line) and with $\mathbf{J}_{RE}=\mathbf{J}$ (red lines).}
\end{figure}

\cref{fig:linear-growthrate} shows the kink mode growth rates and the real frequencies, for cases of $\mathbf{J}_{RE}=0$ and $\mathbf{J}_{RE}=\mathbf{J}$. The value of the normalized resistivity $\hat{\eta}=\eta R/\left(\mu_0 v_A a^2\right)$ ($R$ and $a$ are major and minor radii, $\mu_0$ is the vacuum permeability) is varied. Note that $\hat{\eta}$ can be regarded as the inverse of the Lundquist number $S$. The results show that both the growth rate and the real frequency follow the 3/5 power law of $\hat{\eta}$. Only in the cases with \gls{re} current does the mode have a real frequency, which is consistent with the theoretical analysis in \cite{liu_structure_2020}. The largest resistivity shown in the figure ($\hat{\eta}=3\times 10^{-4}$) is close to the resistivity in DIII-D experiment with plasma temperature $T_e\approx 2$eV. For this case, the mode growth rate is about $0.053 \tau_A^{-1}$ with \gls{re} current.

\begin{figure}[h]
	\begin{center}
		\includegraphics[width=0.45\linewidth]{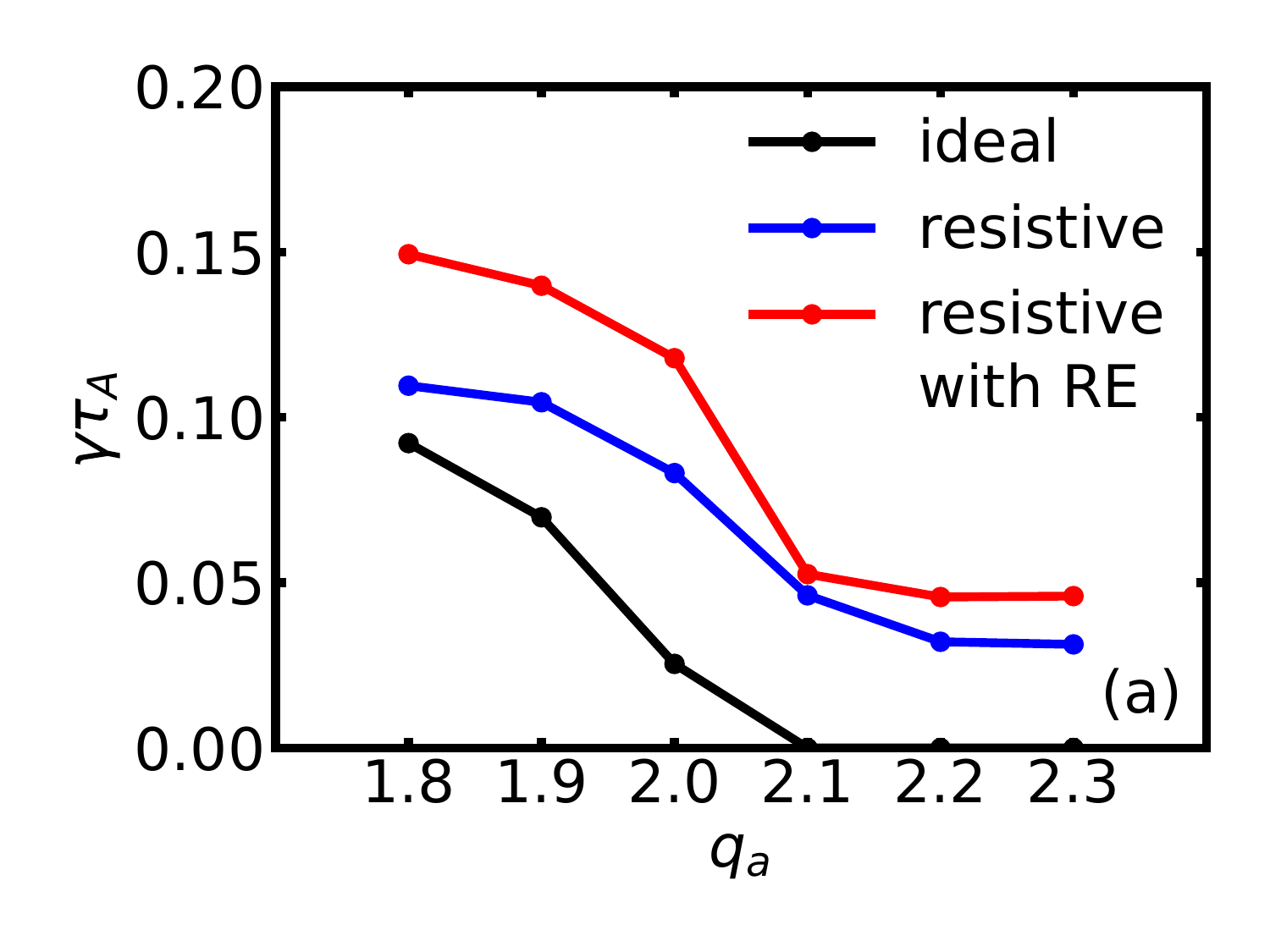}
		\includegraphics[width=0.45\linewidth]{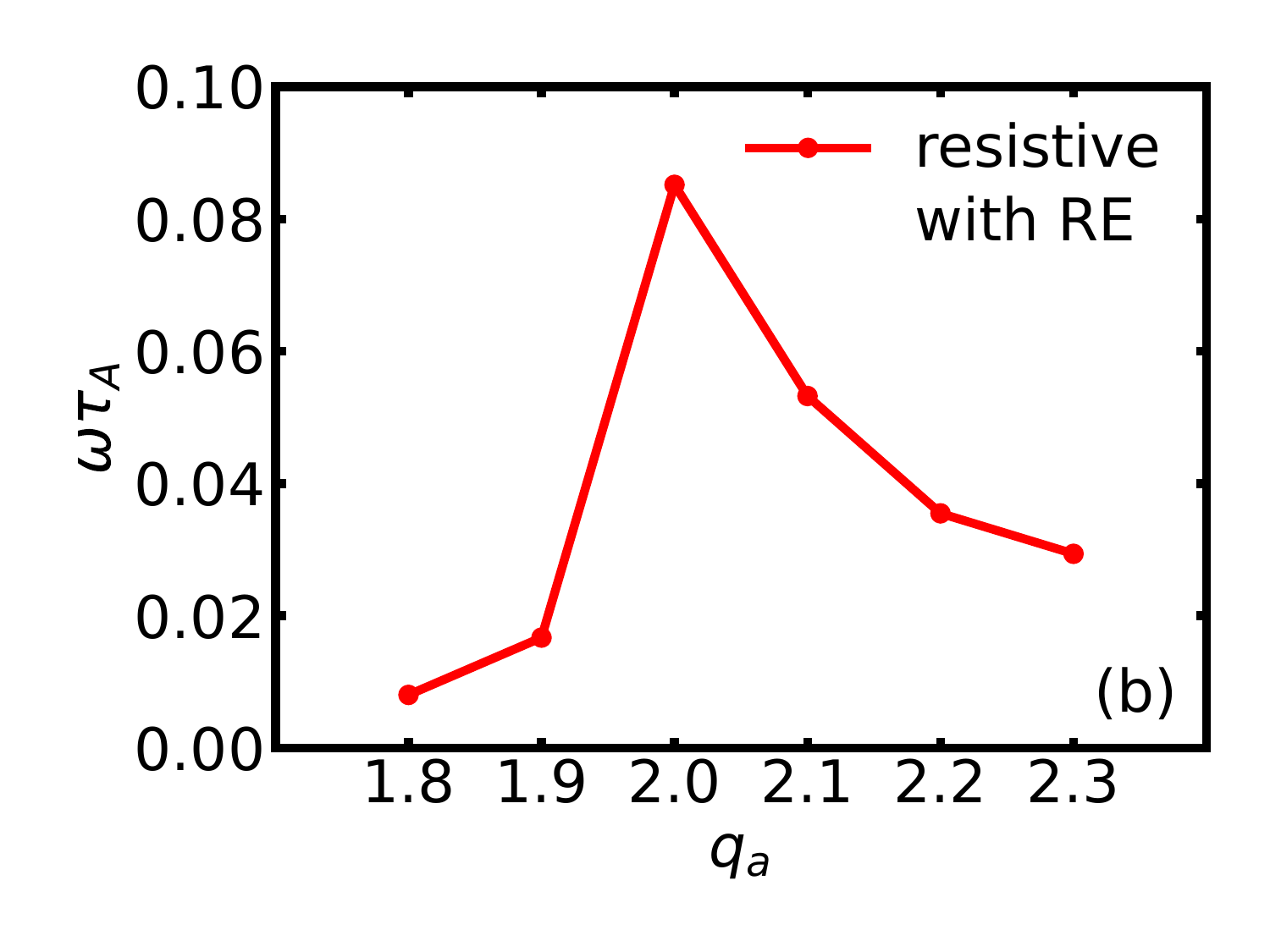}
	\end{center}
	\caption{\label{fig:linear-growthrate-qa}Growth rate $\gamma$ (a) and real frequency $\omega$ (b) of kink mode with different values of $q_a$, for simulations with no resistivity inside \gls{lcfs} (black), with a large resistivity (blue), and  with both resistivity and RE current (red). $\hat{\eta}$ inside \gls{lcfs} is set to be $3\times 10^{-4}$ for the resistive simulations. $\hat{\eta}$ outside \gls{lcfs} is set to be 1 for all the simulations.}
\end{figure}

In addition to resistivity, we also vary the value of $q_a$ and study the mode growth rate and real frequency. In order to change $q_a$, the value of toroidal field in the equilibrium is scaled while the current density is fixed \cite{bateman_mhd_1978}. In addition to the resistive kink mode, we also study the stability of the ideal kink mode by setting $\hat{\eta}$ inside \gls{lcfs} to be zero, while $\hat{\eta}$ outside is set to be 1 to simulate the vacuum region \cite{ferraro_ideal_2010}. The results are shown in \cref{fig:linear-growthrate-qa}. It is found that the ideal mode becomes unstable for $q_a\le 2$ with the growth rate increasing significantly as $q_a$ drops, which is consistent with the theory \cite{cheng_mhd_1987}. The results of the resistive mode simulations with and without \gls{re} follow this trend, but in those cases the mode is still unstable for $q_a>2$ due to large plasma resistivity. The real frequency of the kink mode with \gls{re} increases as $q_a$ decreases when $q_a\ge 2$, which is similar to the growth rate. However, when $q_a$ drops below 2, the real frequency drops significantly while the growth rate increases, as shown in \cref{fig:linear-growthrate-qa} (b).

\begin{figure}[h]
	\begin{center}
		\includegraphics[width=0.45\linewidth]{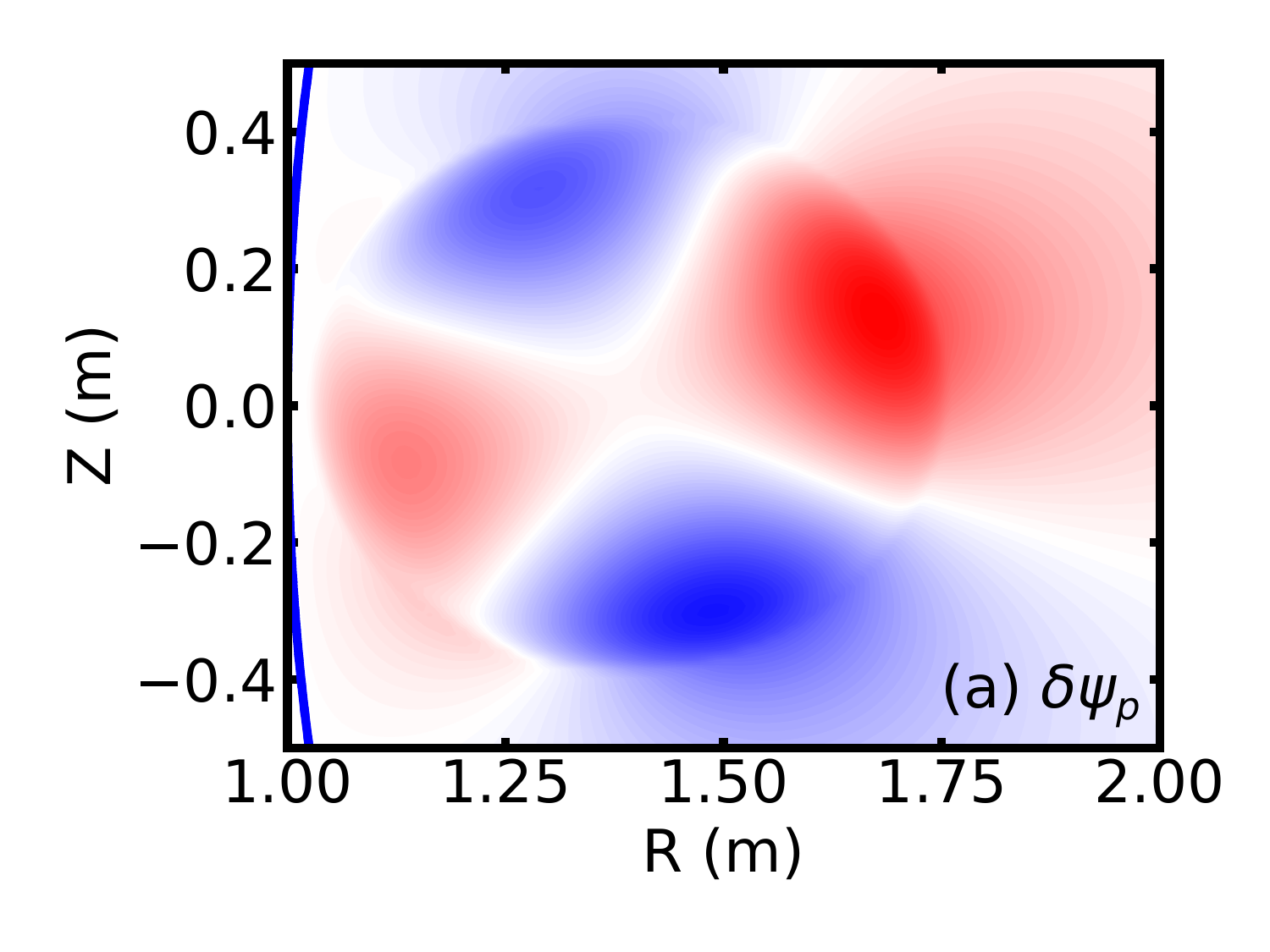}
		\includegraphics[width=0.45\linewidth]{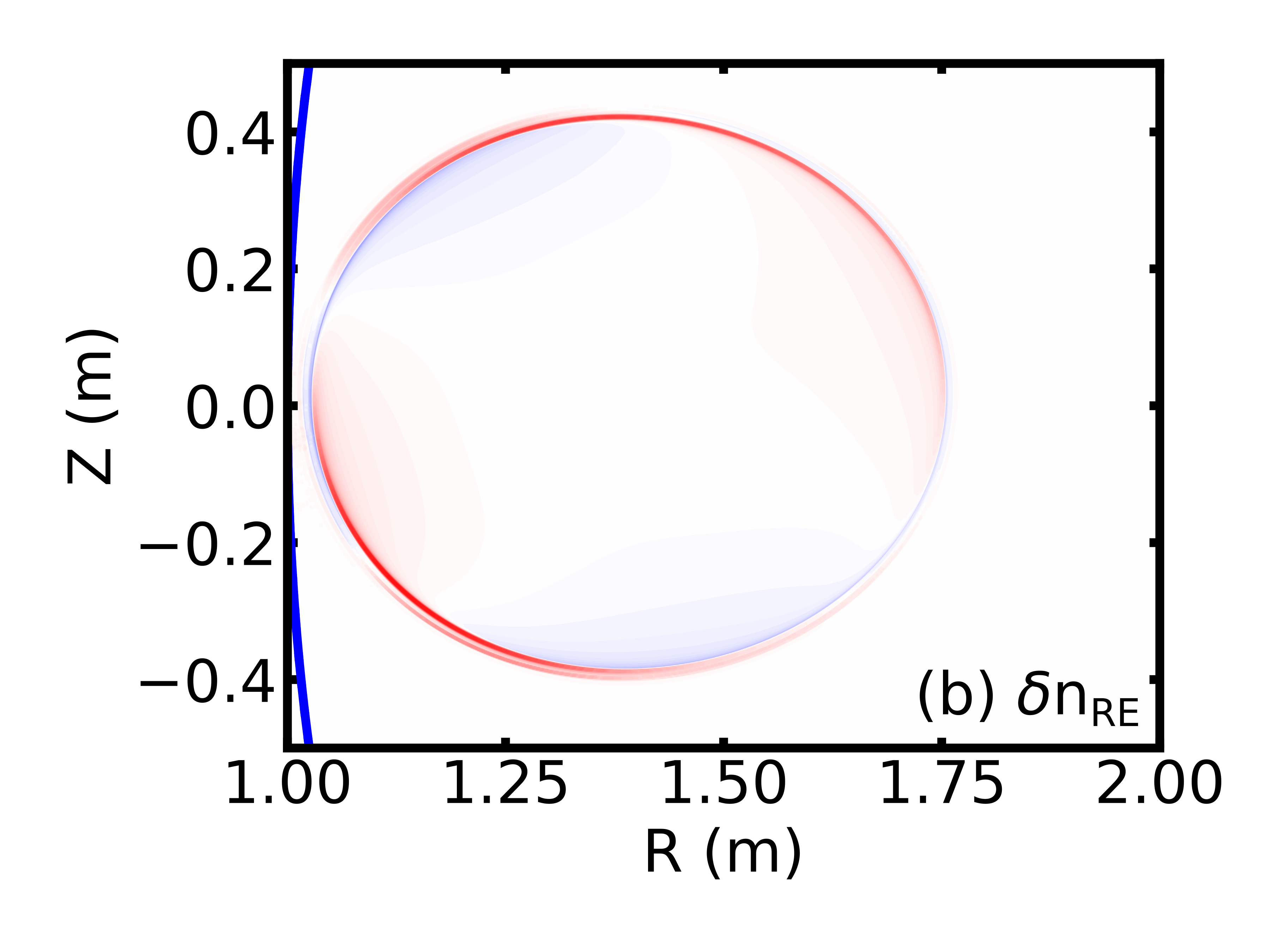}
		\raisebox{0.29\height}{\includegraphics[width=0.05\linewidth]{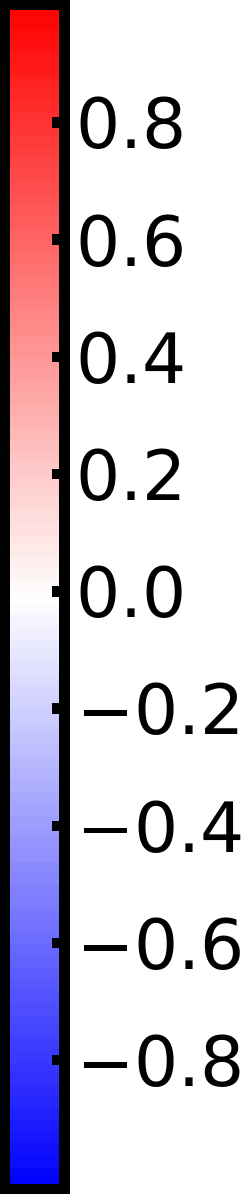}}
	\end{center}
	\caption{\label{fig:delta-psi}Structure of perturbed poloidal flux $\delta \psi_p$ (a) and perturbed \gls{re} density (b) from the linear simulation. The values are normalized according to the maximum absolution value.}
\end{figure}

\begin{figure}[h]
	\begin{center}
		\includegraphics[width=0.5\linewidth]{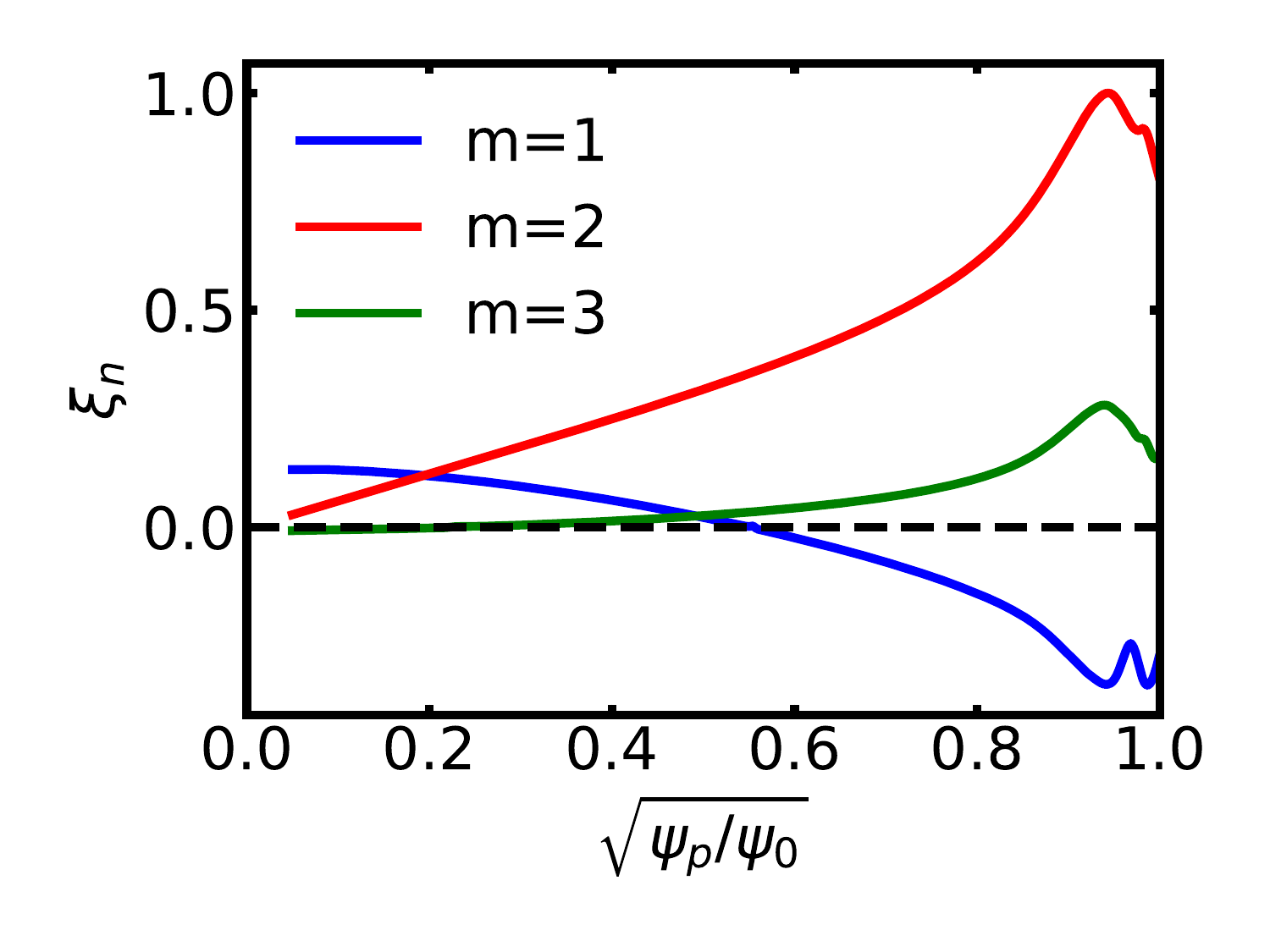}
	\end{center}
	\caption{\label{fig:xin} Radial structure of normal plasma displacement of different poloidal harmonics in the linear simulation.}
\end{figure}

\cref{fig:delta-psi} shows the structure of the kink mode, including the perturbed magnetic poloidal flux ($\delta\psi$) and the perturbed \gls{re} density ($\delta n_{RE}$) for cases with $\hat{\eta}=3\times 10^{-4}$ and $\mathbf{J}_{RE}=\mathbf{J}$. It is shown that the mode is dominated by the $m=2$ component. The perturbed \gls{re} density is localized near the resonant $q=2$ surface.  \cref{fig:xin} shows the radial structure of plasma displacement on the direction normal to flux surfaces ($\xi_n$) decomposed into different poloidal harmonics. The displacement is also localized near $q=2$, which is in agreement with the linear results found with MARS-F \cite{liu_mars-f_2019}.

\begin{figure}[h]
	\begin{center}
		\includegraphics[width=0.5\linewidth]{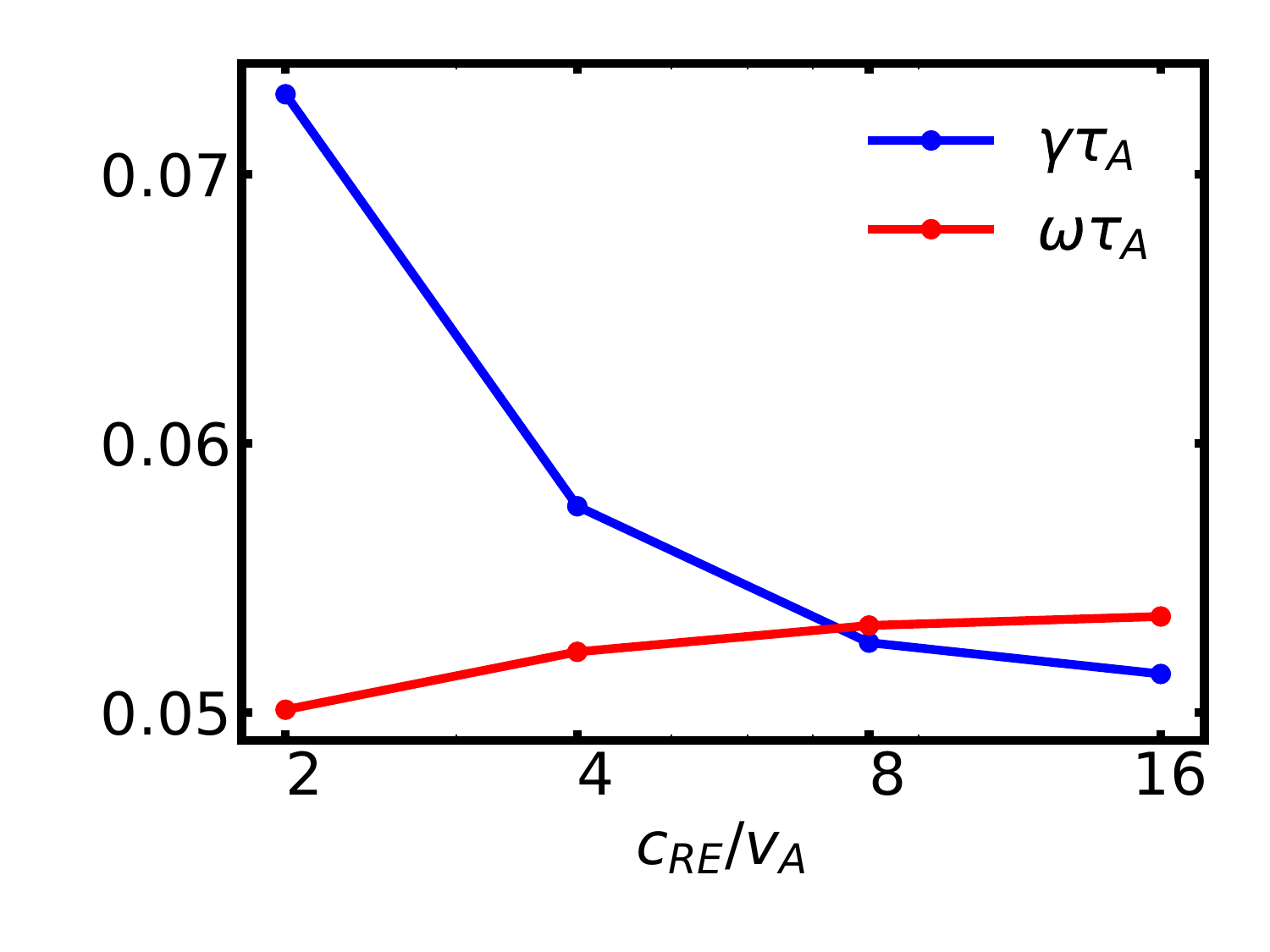}
	\end{center}
	\caption{\label{fig:convergence1}Linear simulation results of mode growth rates (blue line) and real frequencies (red line) using different values of $c_{RE}/v_A$.}
\end{figure}

We did a sensitivity study of the effect of the value of $c_{RE}/v_A$ on the linear growth rates and real frequencies. The result is shown in \cref{fig:convergence1}, which indicates that both $\gamma$ and $\omega$ are not sensitive to this ratio as long as $c_{RE}/v_A>4$. This is also consistent with the analytical theory \cite{liu_structure_2020}.
Similar studies have been conducted regarding mesh density and value of timesteps of \gls{mhd} equation and pseudo particle pushing, and good convergent results are achieved.

\subsection{Nonlinear simulation including \gls{re} loss}
\label{sec:nonlinear-simulation}

Based on the linear simulation result, we did a nonlinear simulation of resistive kink instabilities in 3D, with 16 planes in the toroidal direction. Each plane has structure like in \cref{fig:RE-density-2d} (b) and are connected by Hermite cubic elements in the toroidal direction. In the nonlinear simulation the \gls{re} characteristic line is calculated following the sum of equilibrium and perturbed magnetic fields, and the $\mathbf{E}\times\mathbf{B}$ drift is included. When calculating characteristic lines of \glsps{re} following \cref{eq:characteristic-line}, if the line crosses the mesh boundary, it means that at the position where the pseudo particles originates there is no new \glsps{re} from other locations to replenish \gls{re} density. Thus the value of $n_{RE}$ at this original location is set to zero, which indicates \gls{re} loss to the wall. This  method is equivalent to using an absorbing boundary condition for $n_RE$  when solving \cref{eq:re-convection} directly \cite{zhao_simulation_2020}.

In nonlinear simulations we use non-uniform Spitzer resistivity, which is calculated from the local electron temperature $T_e$,
\begin{equation}
	\eta=0.51\frac{4\sqrt{2\pi}}{3}\frac{e^2 m_e^{1/2}\ln\Lambda}{(4\pi\epsilon_0)^2 \left(k_B T_e\right)^{3/2}}
\end{equation}
where we used $Z_{\mathrm{eff}}=1$. The electron temperature is evolved following \cref{eq:temperature} and \cref{eq:heatflux}, which is controlled by the balance of Ohmic heating and thermal conduction. In this simulation, large thermal conduction (both $\kappa_\parallel$ and $\kappa_\perp$) is set to represent strong collisional diffusion in a post-disruption plasma. Before the final loss event, due to the strong thermal conduction and the absence of Ohmic current, $T_e$ will drop very quickly to the minimum value (set to be 2eV and controlled by the boundary condition of $T_e$), , which gives a resistivity value $\eta\approx 300 \mu\Omega $m. After the \gls{re} current gets lost in the resistive kink instability, the Ohmic heating will increase $T_e$ thus decreasing $\eta$ in the plasma region.

\begin{figure}[h]
	\begin{center}
		\includegraphics[width=0.45\linewidth]{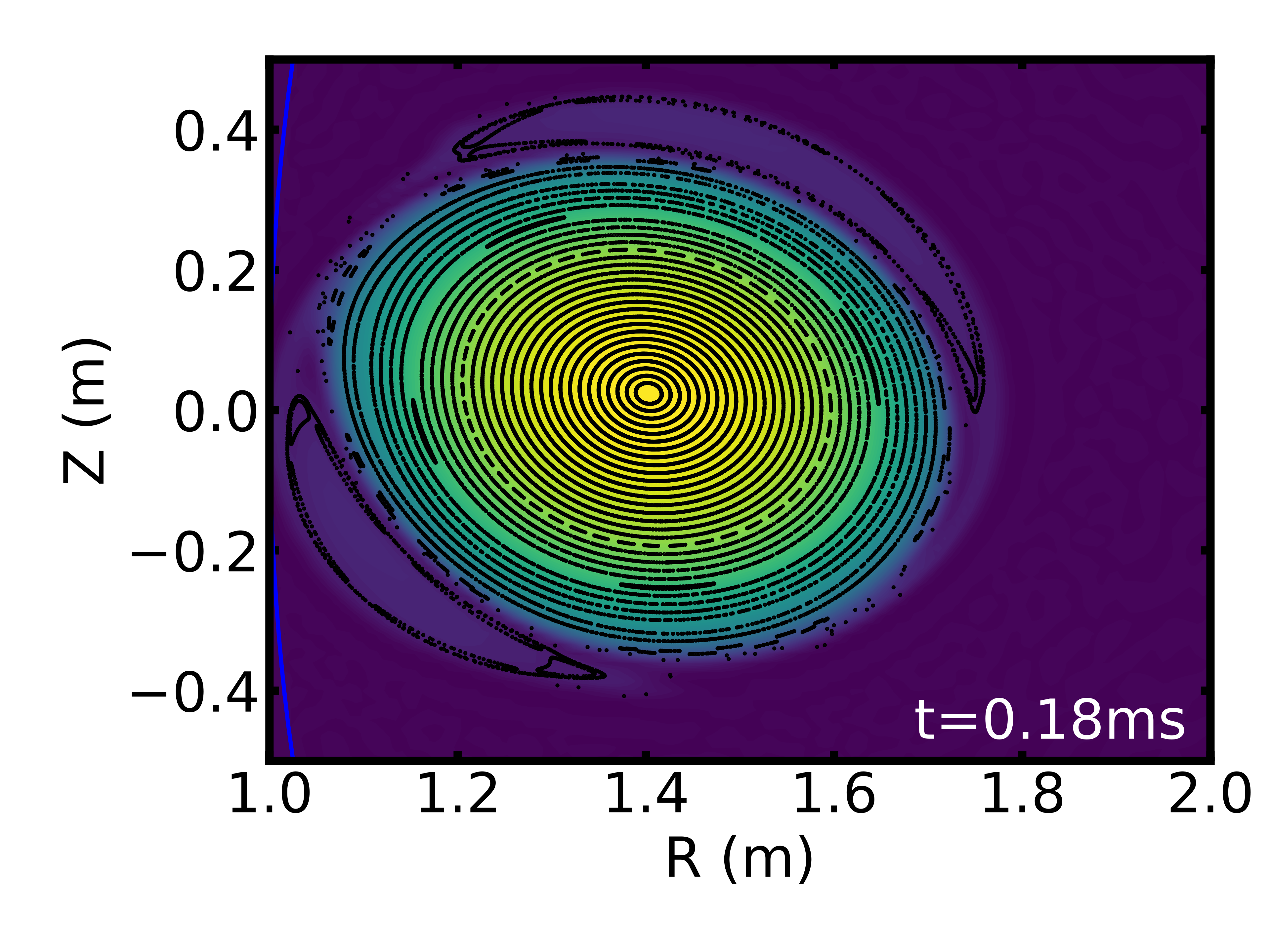}
		\includegraphics[width=0.45\linewidth]{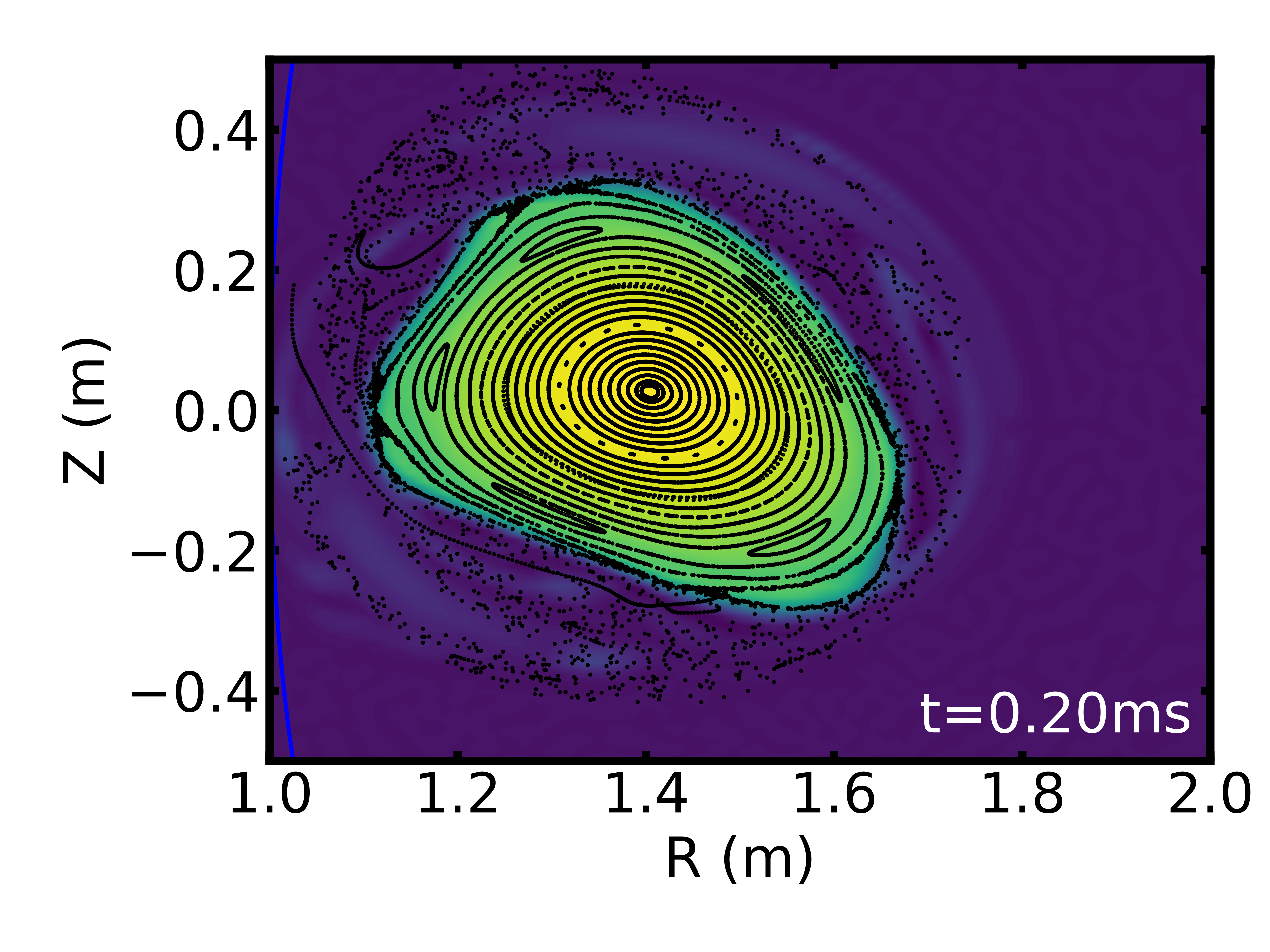}
		\hspace{-1.5cm}\raisebox{0.22\height}{\includegraphics[width=0.18\linewidth]{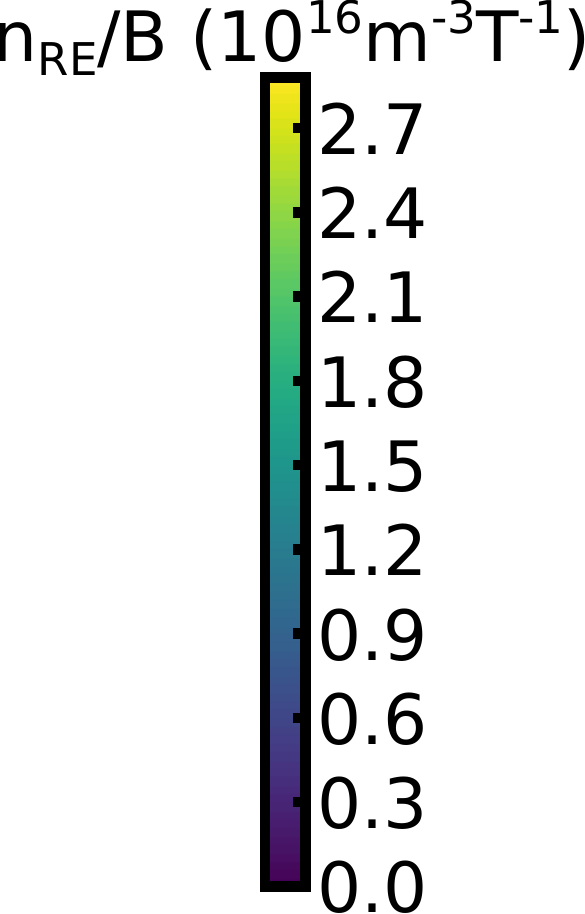}}\\
		\includegraphics[width=0.45\linewidth]{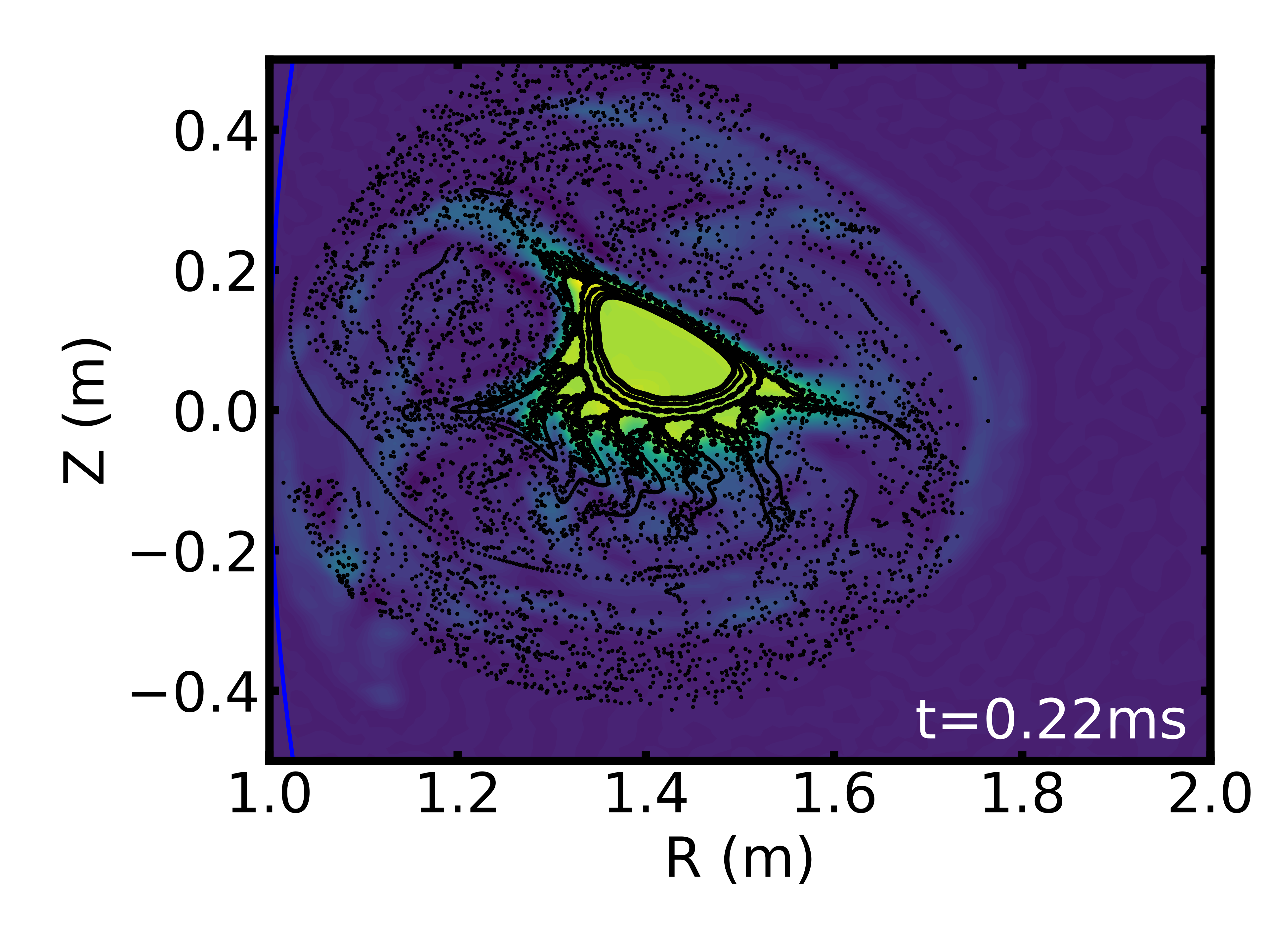}
		\includegraphics[width=0.45\linewidth]{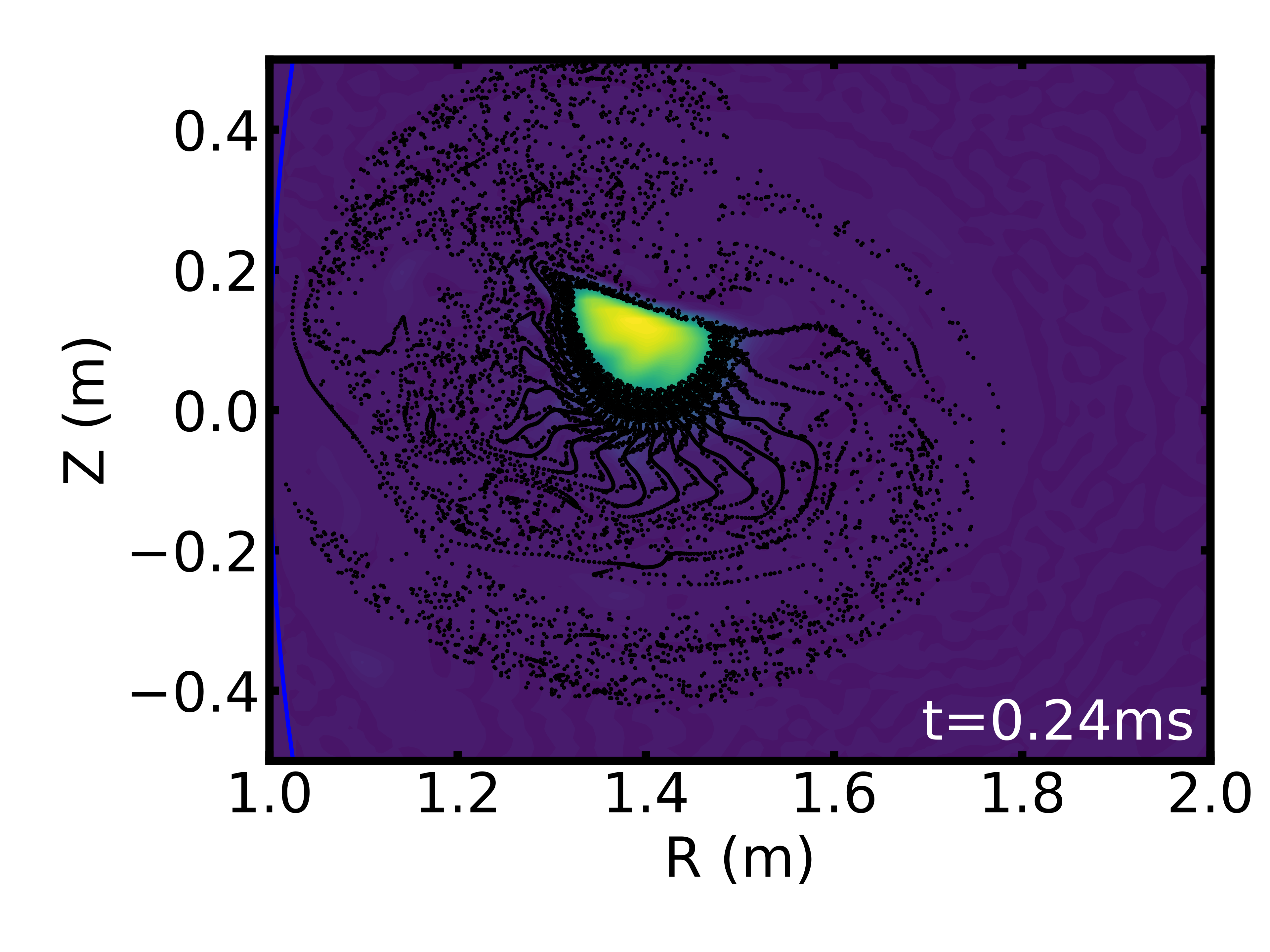}
		\hspace{-1.5cm}\raisebox{0.22\height}{\includegraphics[width=0.18\linewidth]{colorbar2.pdf}}
	\end{center}
	\caption{\label{fig:re-poincare}Evolution of \gls{re} density and Poincaré plots of magnetic field line structure during the nonlinear simulation, at different time.}
\end{figure}

At the beginning of a nonlinear simulation, the (2,1) resistive tearing mode will experience a linear growing stage, and form magnetic islands near the plasma edge, as shown in \cref{fig:re-poincare}. Since the islands are touching the \gls{lcfs}, \glspl{re} initially inside the islands can get dumped into the open field line region and get lost. The islands will also rotate in the plasma frame, but this rotation is not significant since the linear growth only lasts a short time.

\begin{figure}[h]
	\begin{center}
		\includegraphics[width=0.45\linewidth]{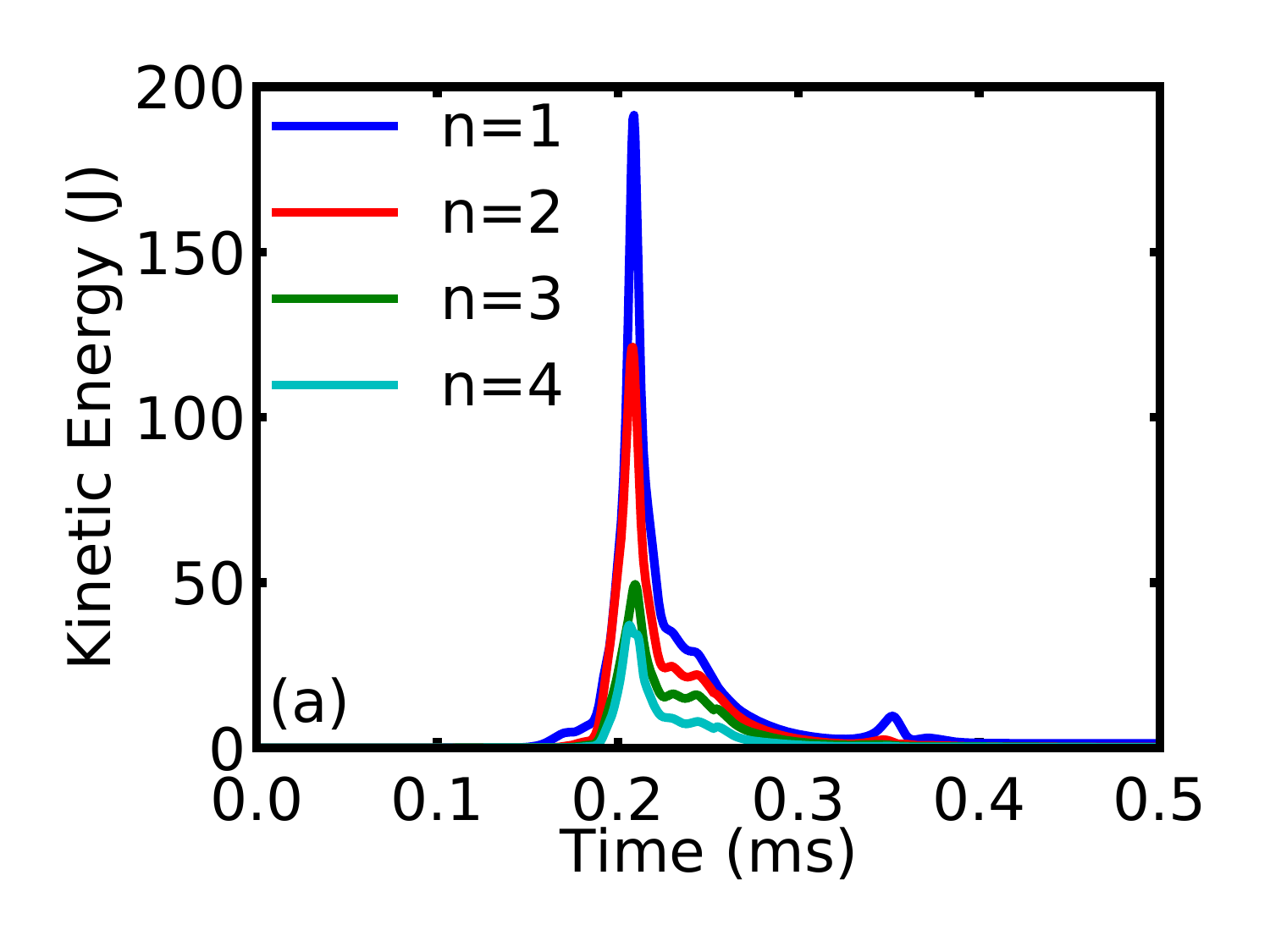}
		\includegraphics[width=0.45\linewidth]{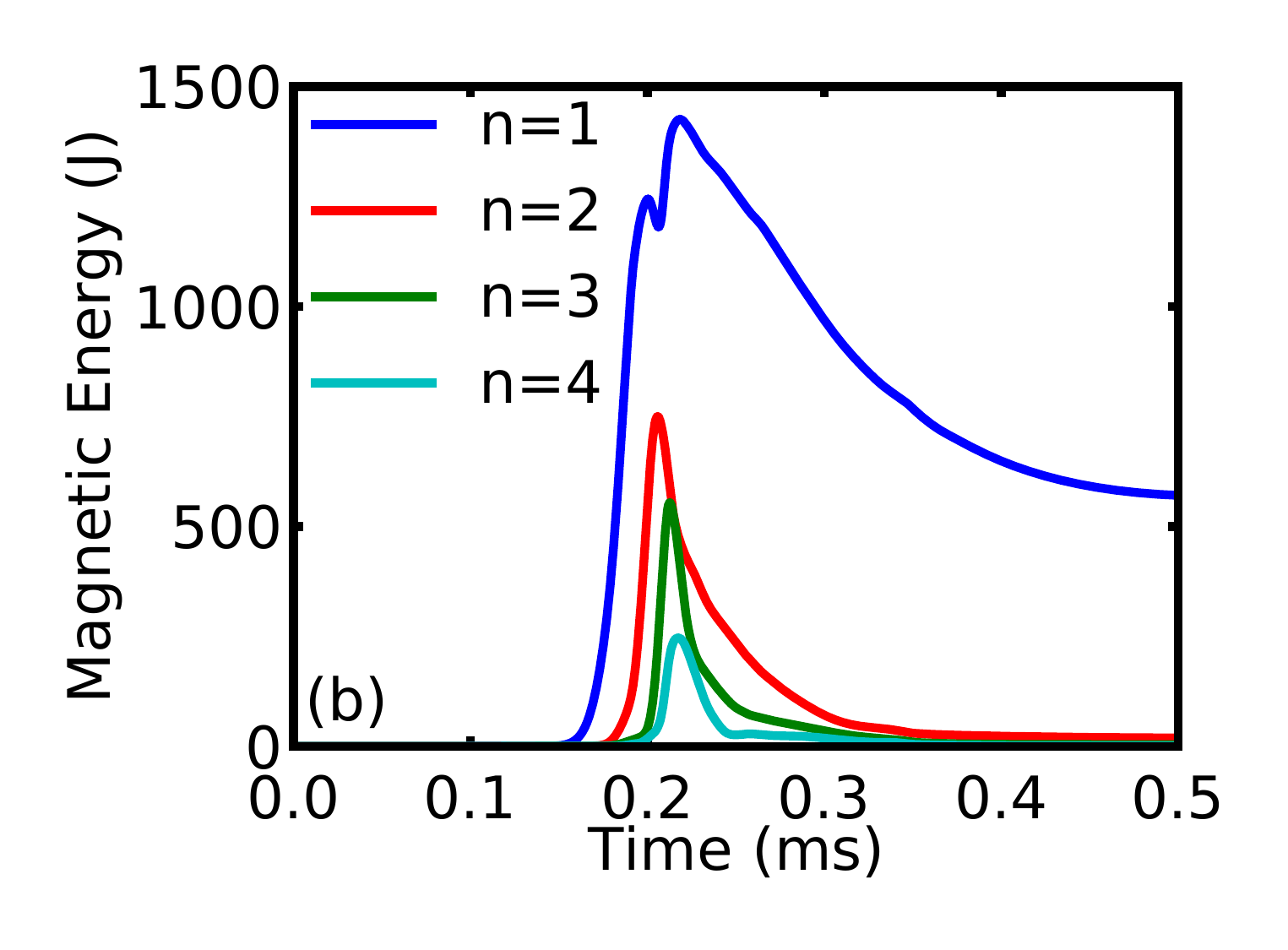}
	\end{center}
	\caption{\label{fig:energy-evolution1}Evolution of kinetic energy (a) and magnetic energy (b) of \gls{mhd} modes with different toroidal mode number in the nonlinear simulation.}
\end{figure}

\cref{fig:energy-evolution1} shows the growth and saturation of kinetic and magnetic energy of \gls{mhd} modes in a nonlinear simulation. Note that in addition to the dominant $n=1$ mode, higher $n$ mode can also get excited due to nonlinear interaction. The kinetic energy of all modes and the magnetic energy of $n>1$ modes all show bursting behavior, while the magnetic energy of the $n=1$ mode has a slow decay following its initial excitation.  When the mode amplitude passes a certain threshold, the field lines become stochastic in the outer region first due to island overlapping, as shown in \cref{fig:re-poincare}. \glsps{re} in the stochastic region get lost very quickly and only \glsps{re} inside closed flux surface remain. The stochastic region  then grows and further breaks inner flux surfaces. Within 0.06ms, almost all the \glsps{re} are lost to the wall, and only the \glsps{re} residing near the magnetic axis remain, where there are still closed flux surfaces.

\begin{figure}[h]
	\begin{center}
		\includegraphics[width=0.45\linewidth]{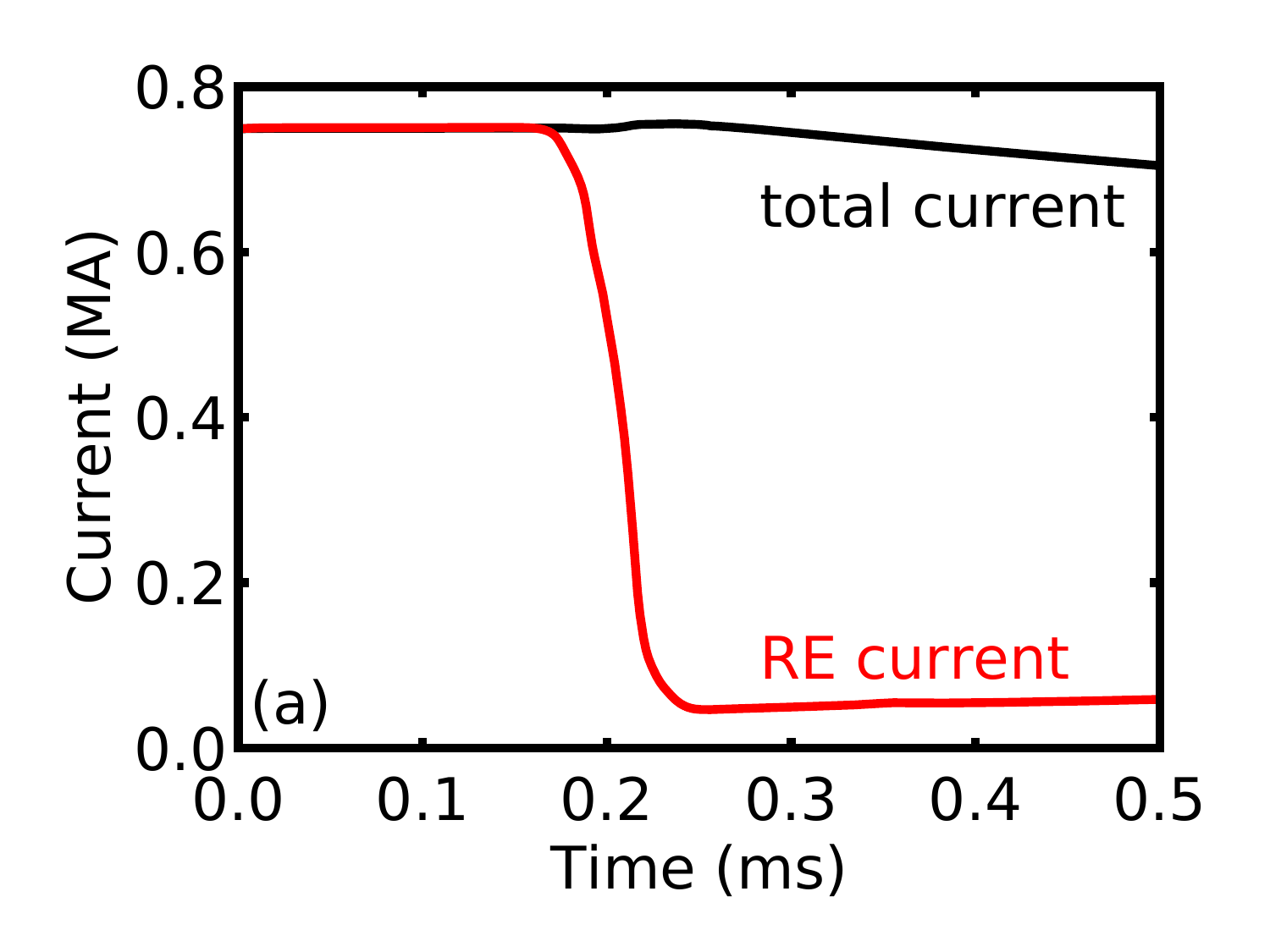}
		\includegraphics[width=0.49\linewidth]{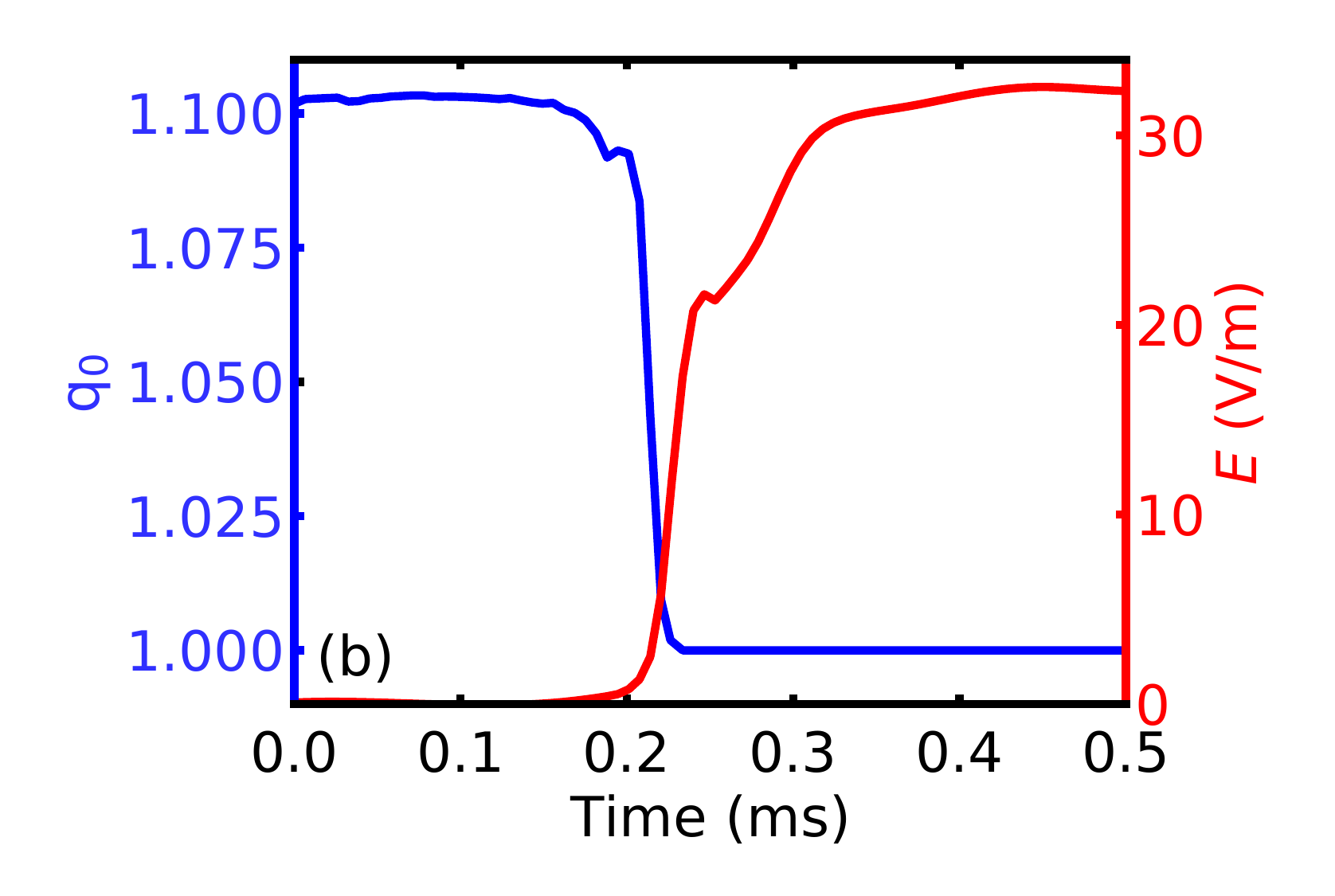}
	\end{center}
	\caption{\label{fig:current-evolution1}(a) Time traces of \gls{re} and total current when resistive kink mode happens. (b) Time evolution of $q$ and electric field at magnetic axis.}
\end{figure}

\cref{fig:current-evolution1} (a) shows the evolution of the \gls{re} current and total current. Note that although the \gls{re} experience significant loss ($>95\%$) during the excitation of resistive kink modes, the total current does not change much. This means that the lost runaway current is replaced by current carried by thermal electrons. This new current has strong collisional resistivity which is balanced by parallel electric field, as shown in \cref{fig:current-evolution1} (b). In addition, the change of magnetic field topology also leads to a flattening of current density in the region of stochastic fields, and the total internal inductance decreases, which further leads to a small current spike of the total current. This current spike was also seen in JOREK simulation \cite{bandaru_magnetohydrodynamic_2021} and has been observed in the DIII-D experiment. The mechanism is similar to the current spike happening during the thermal quench of tokamak disruptions \cite{boozer_runaway_2017}. The inductive electric field can increase the current density at the core region, which can be illustrated by the evolution of safety factor at magnetic axis $q_0$. As $q_0$ drops to 1, a (1,1) kink instability can be excited and cause a flattening of $n_{RE}$ near the core region. 

\begin{figure}[h]
	\begin{center}
		\includegraphics[width=0.43\linewidth]{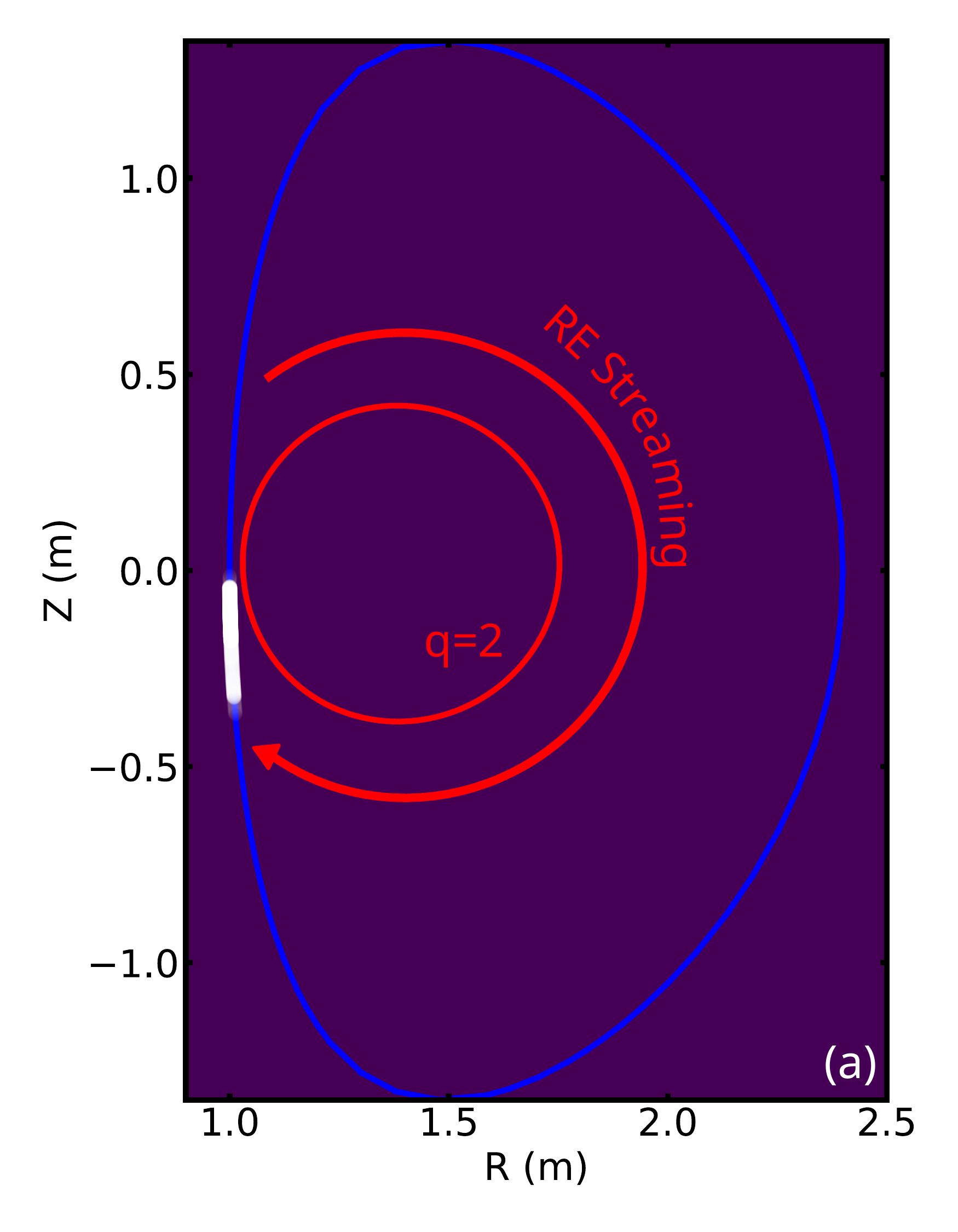}
		\raisebox{0.25\height}{\includegraphics[width=0.45\linewidth]{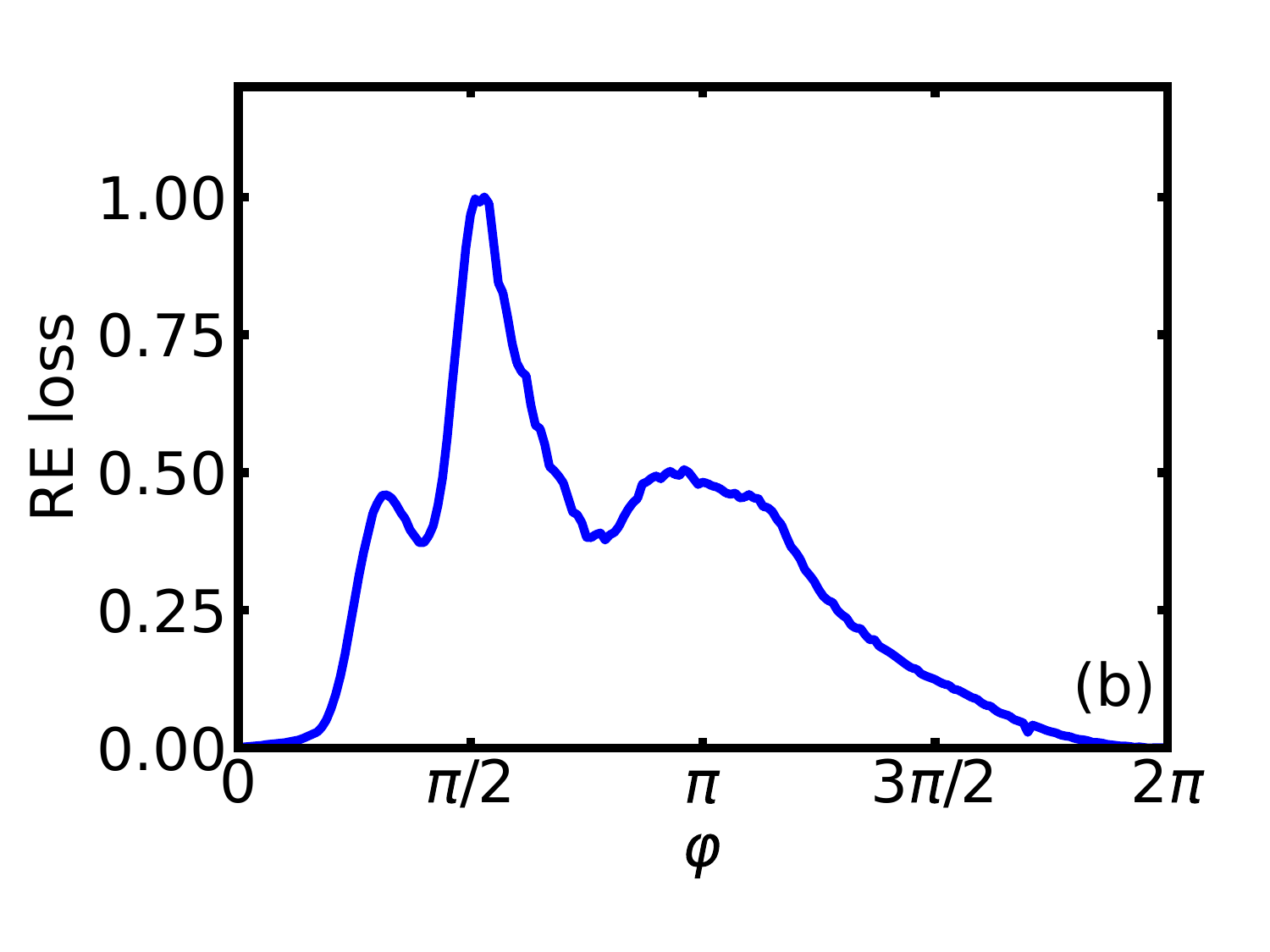}}
	\end{center}
	\caption{\label{fig:spot}Sum of \gls{re} density during kink instability at different location of poloidal boundary (a) and toroidal angle (b). The white spot in (a) represents the peak \gls{re} loss. The red arrow curve indicated the direction of \gls{re} streaming in the simulation.}
\end{figure}

To better understand \gls{re} loss during the instability, we count the total value of $n_{RE}$ lost at different locations of the boundary. \cref{fig:spot} shows the lost \gls{re} deposition location on the poloidal plane (represented by white spots) and toroidal angles. In this discharge, since the \gls{re} beam is streaming in the same direction as the magnetic field and the plasma is only touching the wall on the \gls{hfs}, most of the lost \gls{re} will hit the lower part of the wall on the \gls{hfs} after getting transported into the open field field line region. Since the $n=1$ mode dominates during the \gls{re} loss, the deposition forms a single peak, which is similar to the results of \gls{re} loss simulation in JET using JOREK \cite{bandaru_magnetohydrodynamic_2021}. Note that this analysis is only based on \gls{re} density, as in the fluid \gls{re} model, no \gls{re} energy information is stored.

\begin{figure}[h]
	\begin{center}
		\includegraphics[width=0.45\linewidth]{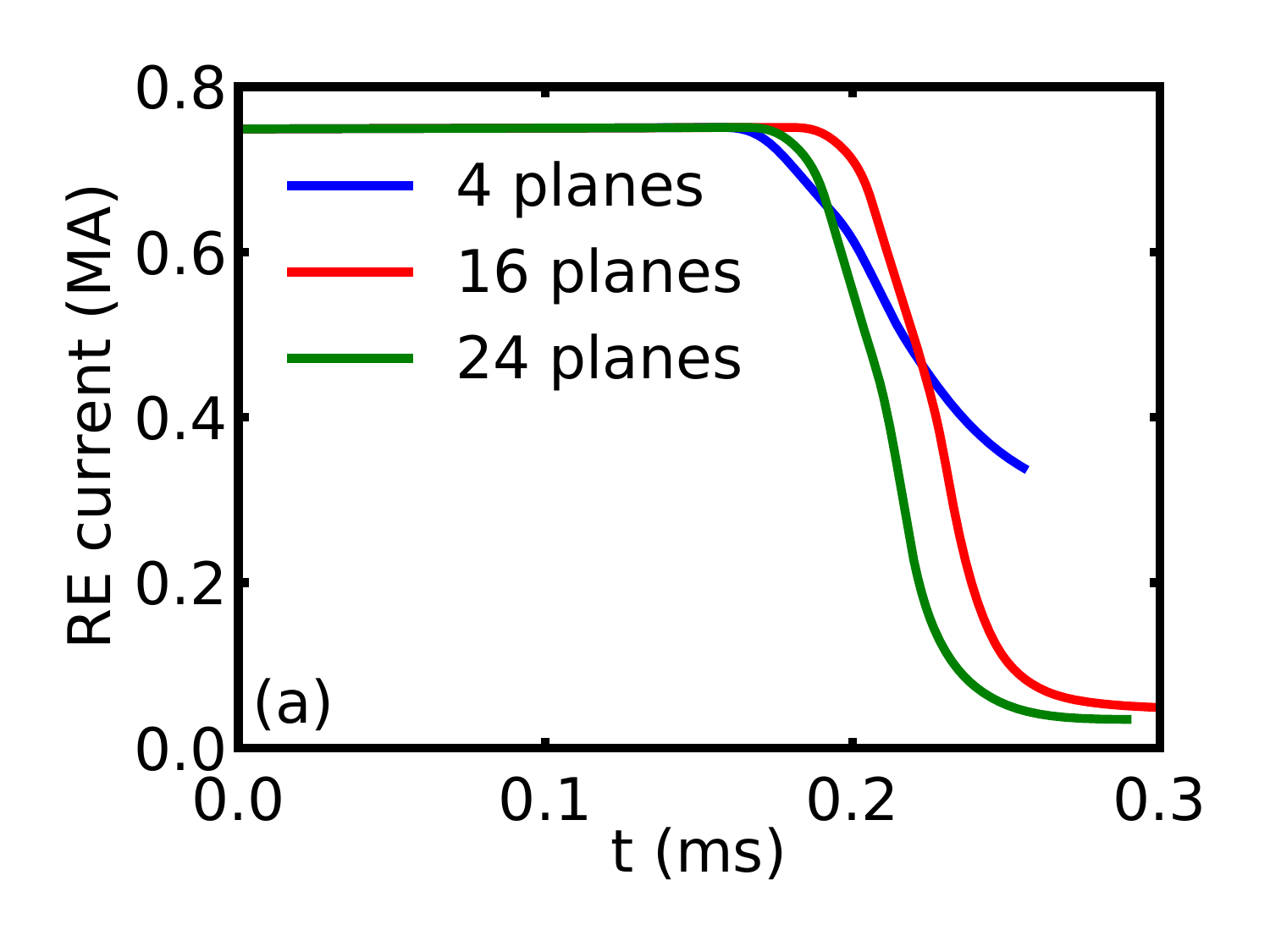}
		\includegraphics[width=0.45\linewidth]{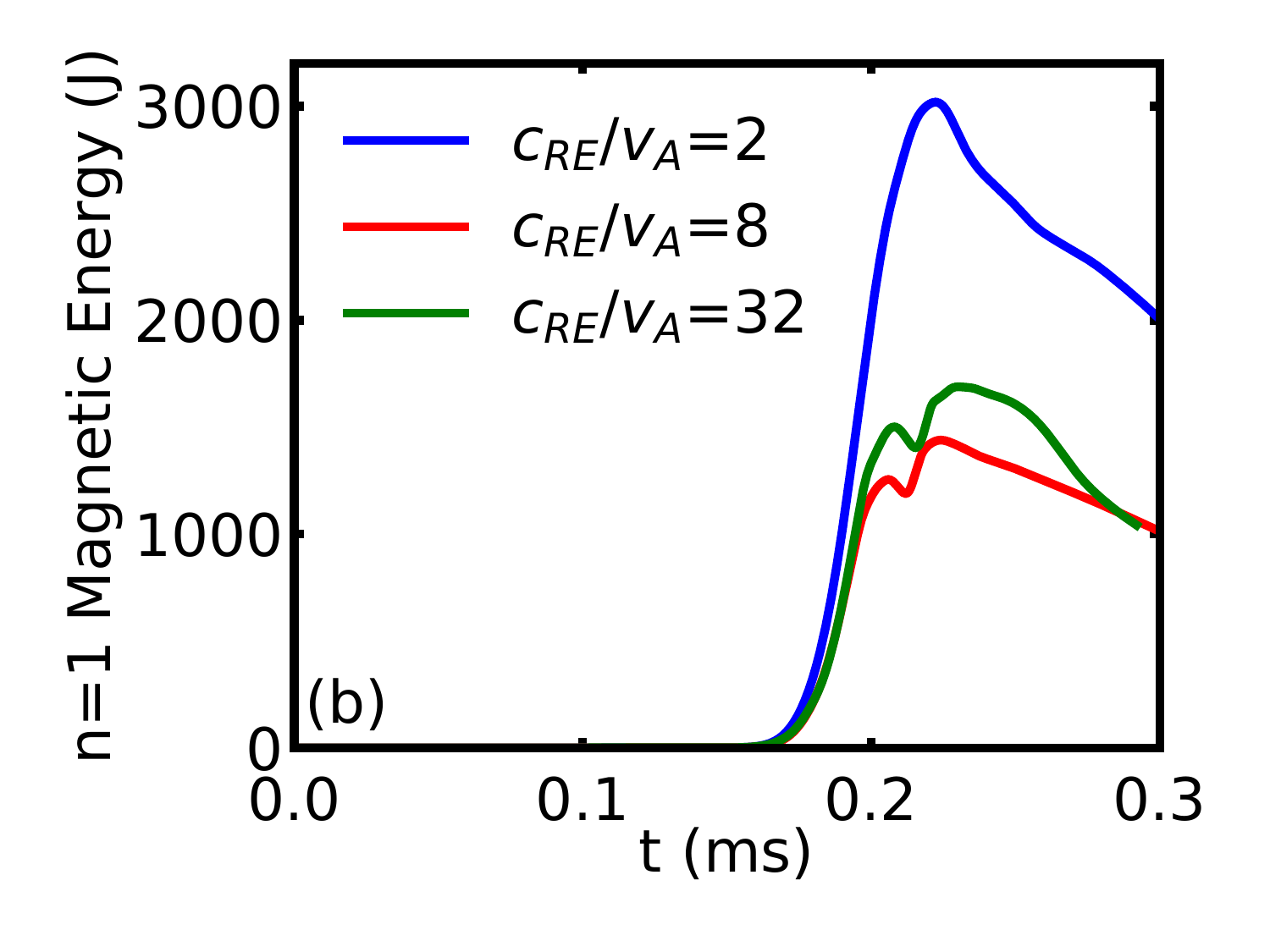}
	\end{center}
	\caption{\label{fig:RE-current-convergence}Convergence study of \gls{re} current loss ratio on different number of toroidal planes (a), and evolution of magnetic energy of $n=1$ mode using different value of $c_{RE}/v_A$.}
\end{figure}

We also did some convergence studies regarding parameters used in the nonlinear simulation, including the number of toroidal planes for the 3D mesh, and the value of $c_{RE}$. The results are shown in \cref{fig:RE-current-convergence}. We found that by increasing the number of toroidal planes, the \gls{re} loss ratio increases, and there is a significant different between 4 planes and 16 planes. In M3D-C1, cubic Hermite polynomials are used in toroidal direction to represent quantities between adjacent planes, so 4 planes is usually enough to accurately represent $n=1$ mode. However, this difference in \gls{re} loss ratio shows that higher $n$ modes play an important role in the formation of stochastic fields and determining \gls{re} loss. In the study of varying $c_{RE}$, it is found that although this change does not significantly affect the \gls{re} loss ratio, it has an impact on the mode saturation level. As shown in previous studies \cite{helander_resistive_2007}, the existence of \gls{re} current plays an important role in determining the mode saturation amplitude. Thus if \gls{re} get lost very quickly due to the large value of $c_{RE}$, the mode can saturate in a lower level due to the absence of \gls{re} current in later time. This study confirmed the necessity of using a large value of $c_{RE}$ for an accurate nonlinear simulation, and thus the importance of applying the novel numerical method introduced in \cref{sec:characteristic-method,sec:boris-algorithm}.

\subsection{Nonlinear simulation after final loss event}
\label{sec:nonlinear-simulation-after}

We continued the nonlinear simulation after the initial \gls{re} loss. Note that in the experiment, after significant \gls{re} loss, the plasma temperature increases due to the Ohmic heating of thermal electron current, and neutrals can get ionized which leads to an increase of plasma density. In our simulation, we have the Ohmic heating term in the temperature equation, and self-consistent evolution of plasma temperature and Spitzer resistivity, but the ionization of neutrals is not included. So we artificially  increase the plasma density by a factor of 10 at $t=$0.3ms, in order to match the increase of plasma density in the experiment. For \gls{re} density we include a source term for secondary generation due to knock-on collisions, using Eq.~(7) from Ref. \cite{rosenbluth_theory_1997}, since in this case the electric field is much larger than Connor-Hastie critical field ($E/E_{CH}\approx 10$) and there are no high-$Z$ impurities.

\begin{figure}[h]
	\begin{center}
		\includegraphics[width=0.45\linewidth]{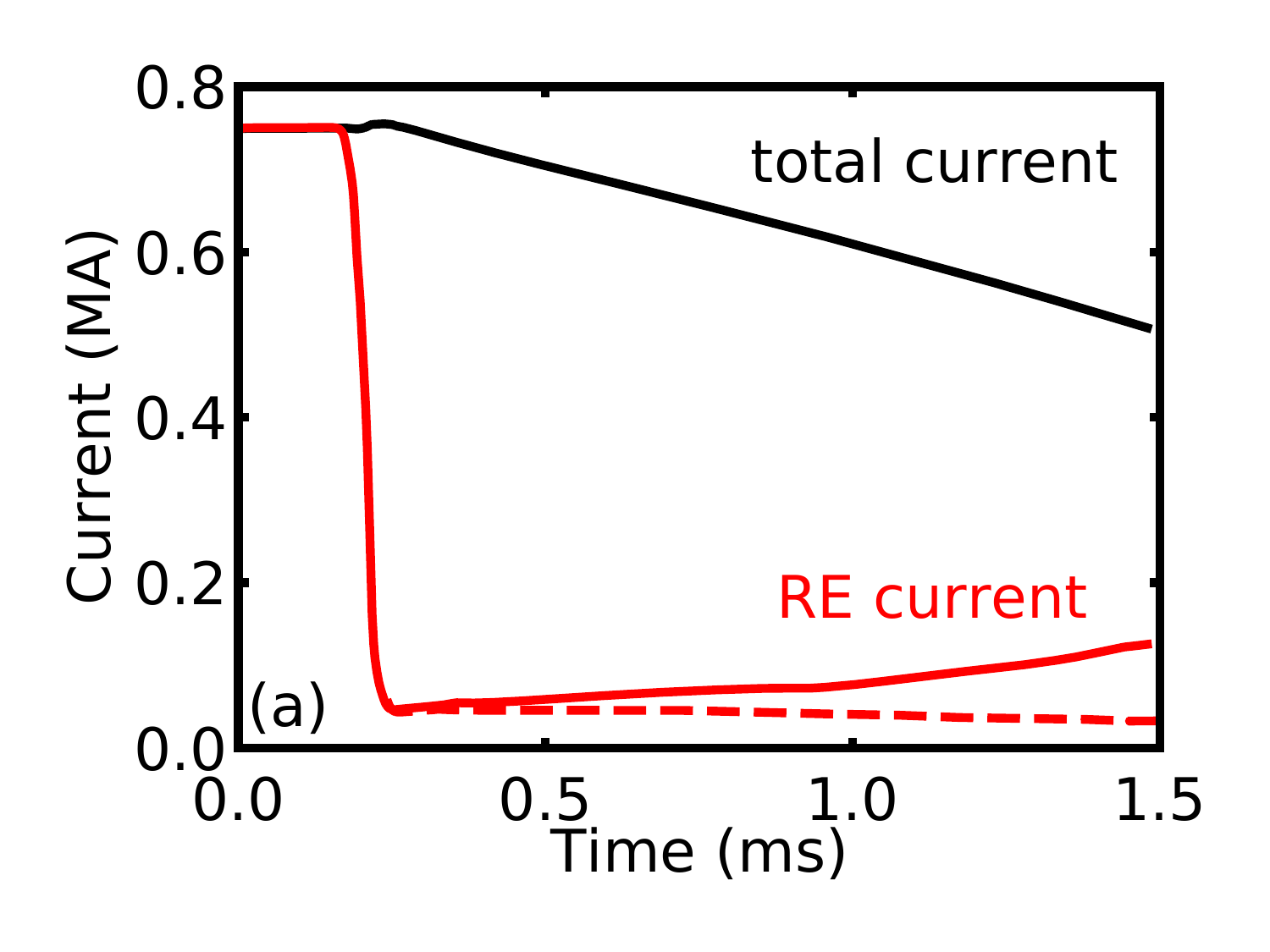}
		\includegraphics[width=0.45\linewidth]{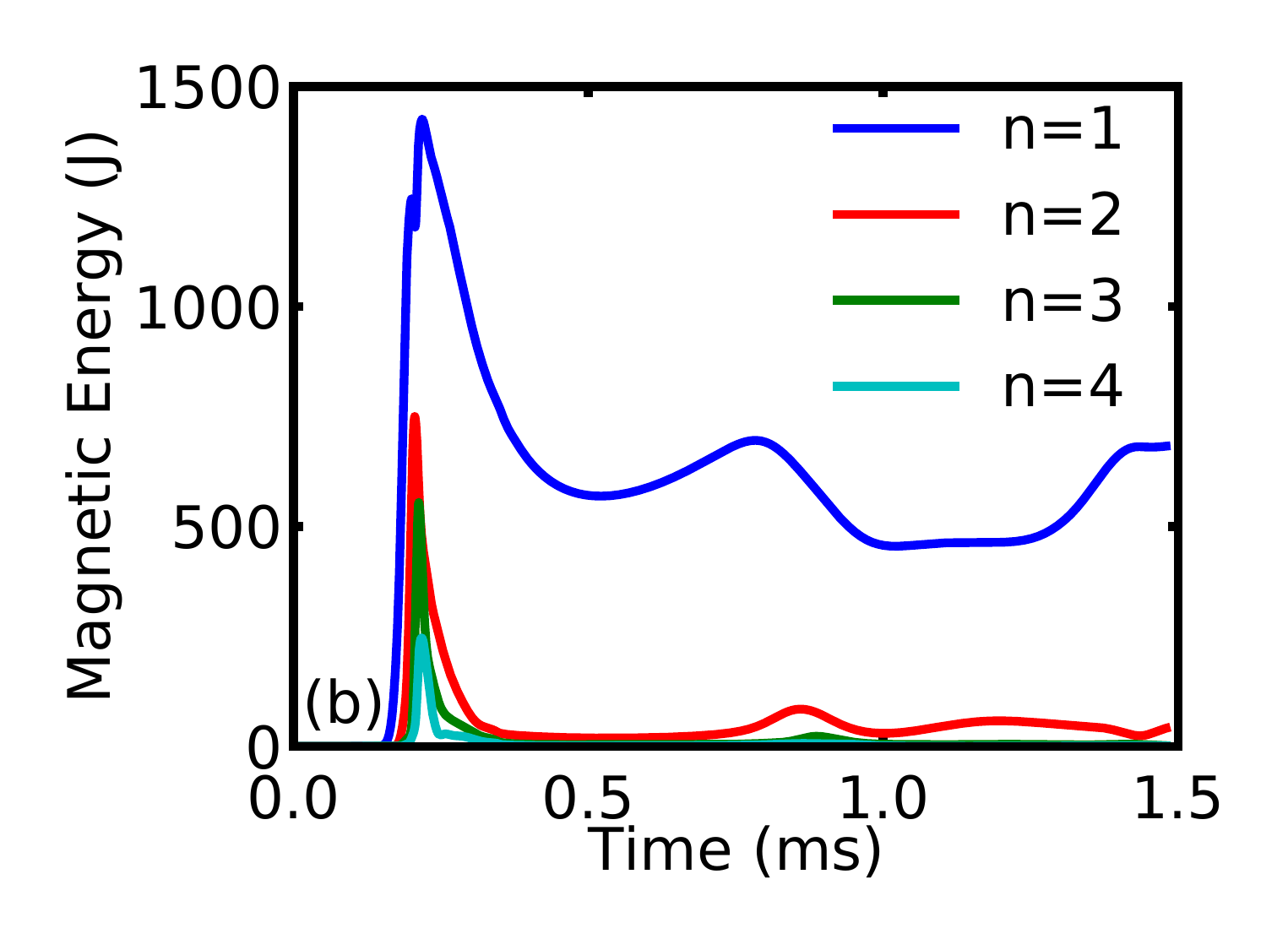}
	\end{center}
	\caption{\label{fig:current-evolution2}(a) Time traces of \gls{re} and total current after kink instability and \gls{re} loss event. Dashed line shows the \gls{re} current without secondary generation. (b) Evolution of magnetic energy of \gls{mhd} modes with different toroidal mode number.}
\end{figure}

\begin{figure}[h]
	\begin{center}
		\includegraphics[width=0.45\linewidth]{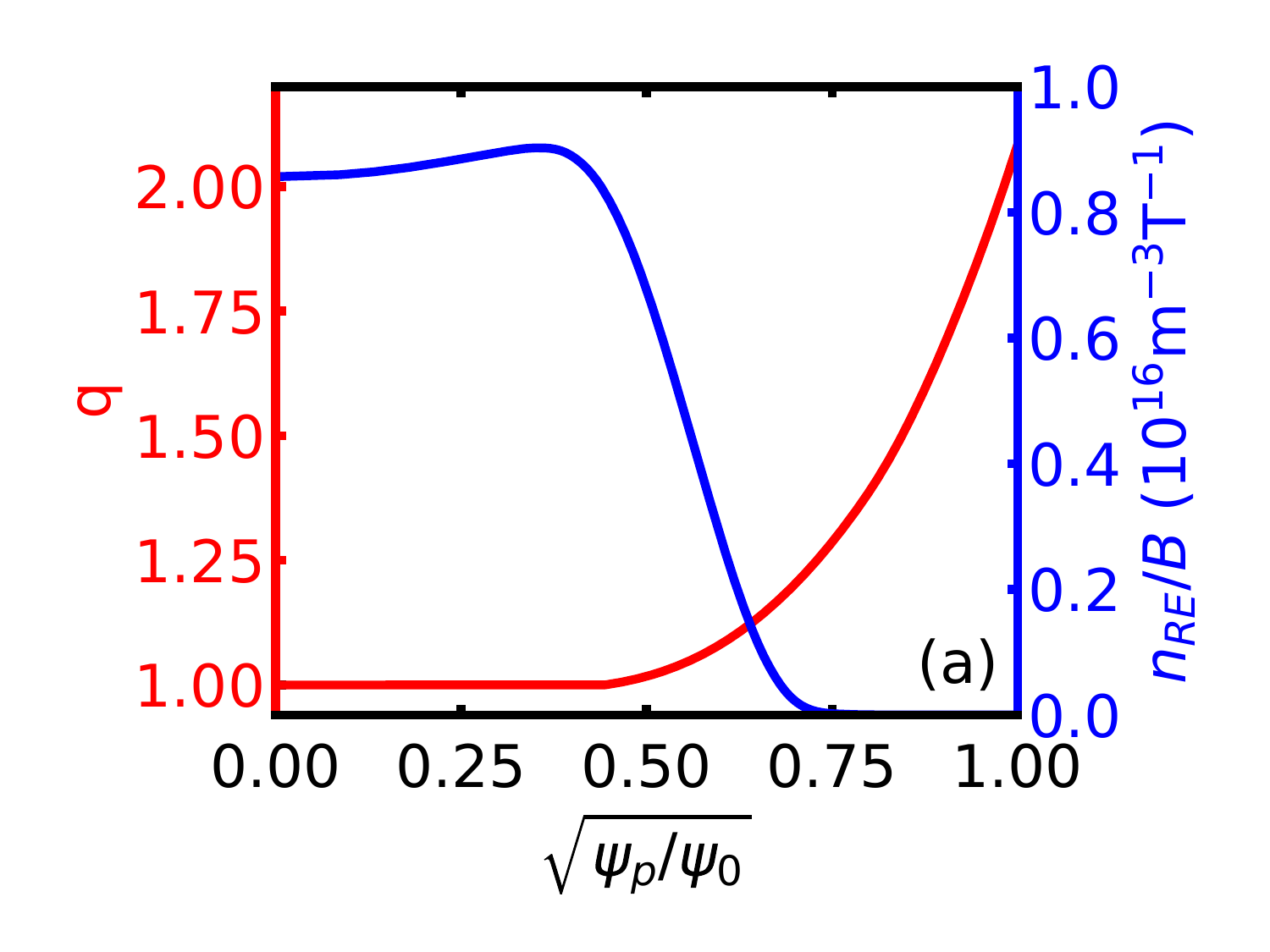}
		\includegraphics[width=0.45\linewidth]{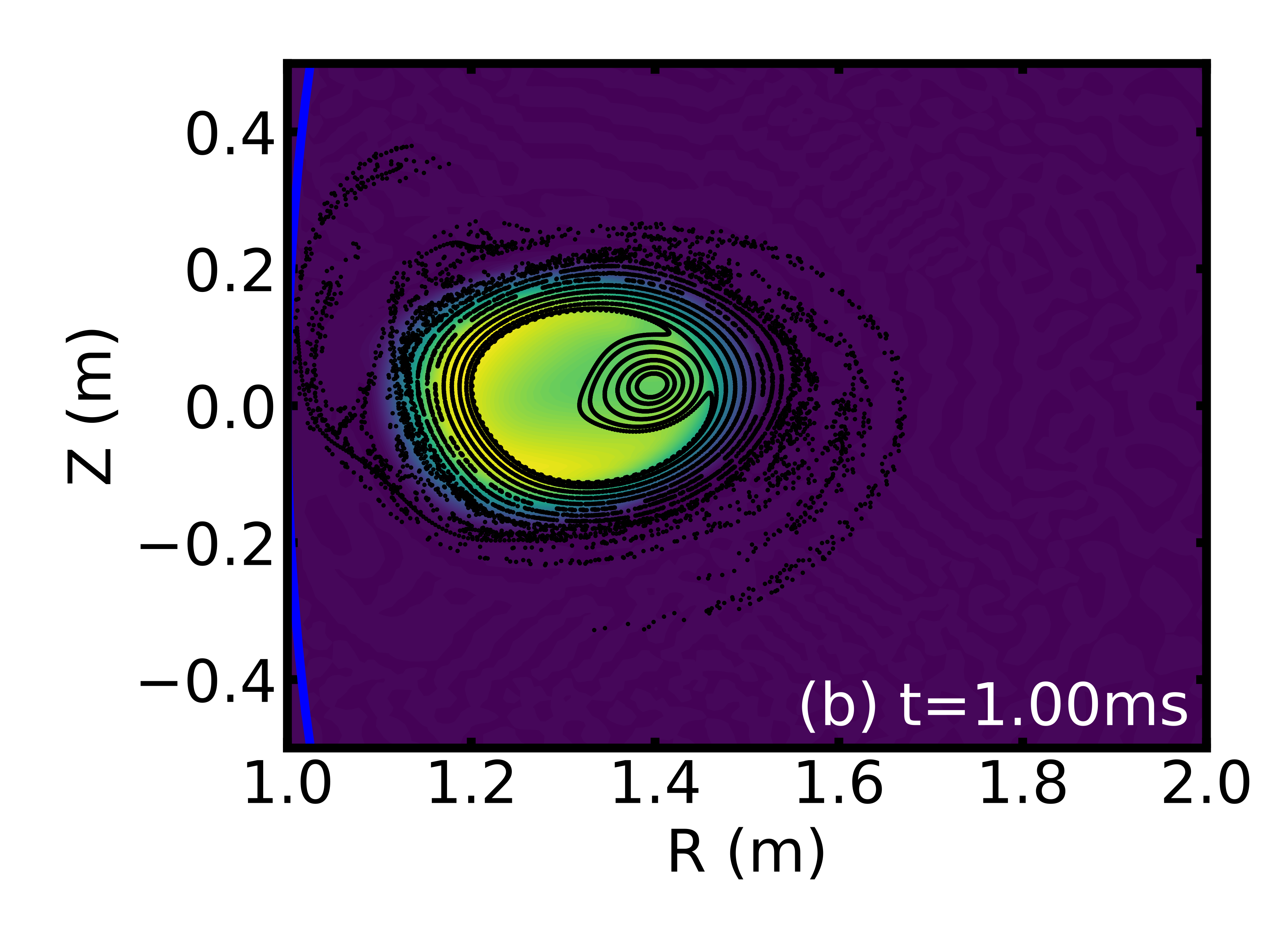}
		\hspace{-1.5cm}\raisebox{0.22\height}{\includegraphics[width=0.18\linewidth]{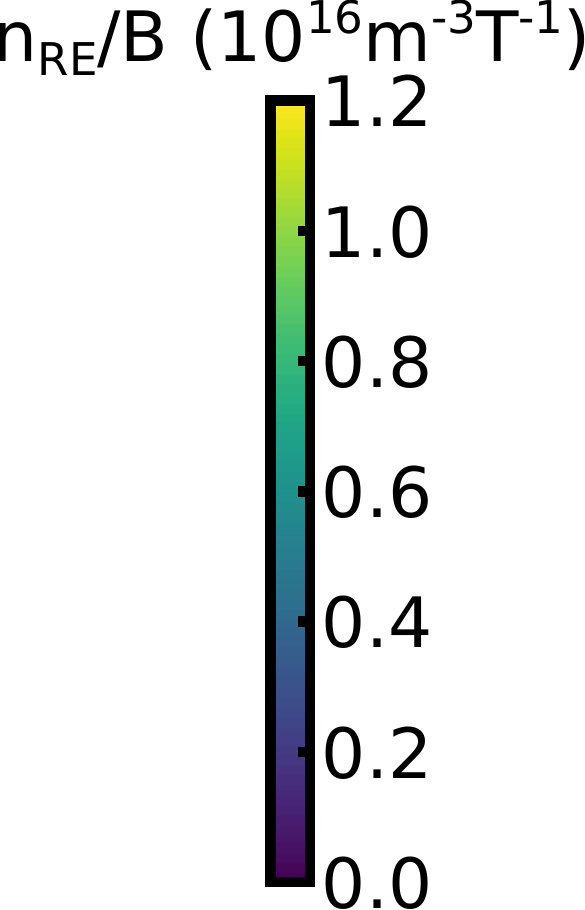}}\\
	\end{center}
	\caption{\label{fig:q-profile2}(a) Profiles of $q$ and $n_{RE}/B$ at $t=1.0$ms. (b) \gls{re} density and Poincaré plot of magnetic field line structure}
\end{figure}

\cref{fig:current-evolution2} (a) shows the evolution of \gls{re} current and total current in the later time. The total plasma current, which mainly consist of Ohmic current, decays due to the resistive diffusion. Ohmic heating causes $T_e$ to rise to about 20eV, which gives a resistivity value $\eta\approx 15 \mu\Omega $m. The current decay time is about 3 ms, which is consistent with experimental observation. The \gls{re} current grows due to the secondary generation, but it is still much smaller compared to the total current. \cref{fig:current-evolution2} (b)  shows the evolution of magnetic energy of \gls{mhd} modes. The $n=1$ mode will remain at a certain level after the initial excitation, and keep the magnetic field stochastic in the outer region. The $n=1$ component includes both (2,1) and (1,1) modes, the latter of which can keep $q_0$ close to 1 during this time, and $q$ and $n_{RE}$ will remain flat near the core region, as shown in \cref{fig:q-profile2}, which is similar to a sawtooth. Since there is no horizontal control in our simulation, plasma and \gls{re} beam will shift toward \gls{hfs} during resistive current decay. Eventually the plasma will hit the wall on \gls{hfs} and all the remaining \gls{re} will get lost.

\section{Summary}
\label{sec:summary}

In this paper we introduced new developments of the fluid model of \glspl{re} in M3D-C1, which helps to enable us to perform self-consistent simulations of \gls{mhd} instabilities with \gls{re} current using realistic physical parameters. In order to simulate the continuity equation of \gls{re} density with large convection speed, we applied the method of characteristics to convert the equation into a form of \gls{ode}, and then solve it by advancing pseudo particles. This method is further optimized using the modified Boris algorithm to push pseudo particle, which can help reduce the accumulation of numerical error.

The newly developed code is used to simulate a resistive (2,1) kink instability happening in a post-disruption \gls{re} plateau in the DIII-D experiment. In the linear simulation, good agreement is obtained between M3D-C1 and MARS-F results, including the growth rate and mode structure. In the nonlinear simulation, it is found that the (2,1) mode can grow to significant amplitude within 0.6ms. Together with other modes, the kink instability can break flux surfaces and make the field lines in the outer region stochastic, which leads to $>95\%$ of \glsps{re} getting lost and only \glsps{re} near the magnetic axis can remain. The plasma current changes from \gls{re} current to Ohmic current, and then exhibits a slow decay due to resistive diffusion. \gls{re} deposition shows $n=1$ toroidal variation and is localized near the \gls{hfs}. After the initial loss event, the strong parallel electric field can cause \gls{re} density increase through secondary generation, and keeps the current density at the magnetic axis high and the safety factor $q$ close to 1.

It is found from the convergence study of nonlinear simulation that high $n$ modes play important roles in determining the loss ratio of \glsps{re}, even though the $n=1$ mode dominates. These high $n$ modes are excited through nonlinear interaction, and can increase the stochasticity of the magnetic field thus accelerating \gls{re} loss. Note that in M3D-C1 we do not use an ad hoc diffusion model for \gls{re} density calculation but a pure convection model, thanks to the method of characteristics. It is found in the linear simulation that the mode growth rate does not have a strong dependence on the value of \gls{re} convection speed $c_{RE}$, which is consistent with the JOREK result \cite{bandaru_magnetohydrodynamic_2021}. For nonlinear simulations, the value of $c_{RE}$ can be important in determining the saturation amplitude of the kink mode, which indicates potential important nonlinear physics associated with \gls{re} convection. The \gls{re} loss ratio though is not sensitive to $c_{RE}$. We thus believe it is important to use more realistic simulation model for \glsps{re} in some cases, and it justifies the necessity of further optimization of the algorithm used in \gls{re} fluid model.

Note that in the current fluid model of \glsps{re}, the gradient and curvature drifts are not included in the \gls{re} continuity equation. However, it is found that the curvature drift can be important for high energy \glsps{re}, and can lead to a shift of \gls{re} drift orbit center from magnetic axis $d=q p_\parallel/\left(e B\right)$ \cite{guan_phase-space_2010}. This shift can lead to a deviation of \gls{re} current iso-surface from magnetic surface, which make it difficult to form an equilibrium satisfying Grad-Shafranov equation. In DIII-D equilibrium, this shift is relatively small due to the small \gls{re} energy, and can be controlled using  external vertical field coils. For JET and ITER, \glsps{re} from disruptions can be more energetic due to better confinement, but the maximum energy is also limited by synchrotron radiation given the larger magnetic field. 

In the simulation model used in the paper, no seed \gls{re} generation mechanism such as Dreicer generation or hot-tail generation is included. This is valid as the plasma temperature remains below 20eV in the whole post-disruption phase, thus the seed generation from the Maxwellian tail is ignorable. For future simulation work targeting thermal and current quench phases at the beginning of the disruption, these seed generation mechanisms should be included, and additional source terms such as tritium decay and Compton scattering should also be included if simulating ITER D-T discharge.

In the experiment, it is observed that \gls{mhd} mode growth and \gls{re} loss can happen in Alfvén timescales. In the simulation the kink instability grows in resistive timescale, but here the difference between the two timescales is small as the Lundquist number in post-disruption plasma is only about $10^3$, which is much smaller compared to that before disruption. Nevertheless, it is possible that fast magnetic reconnection can play a role in this process which can flatten the ratio of $J/B$ along the magnetic field line within Alfvén timescale \cite{boozer_runaway_2017}. One evidence of fast reconnection is that the current spike observed in our nonlinear simulation is still smaller compared to that in  experiment. Simulating fast reconnection on the Alfvén timescale is still a challenging job for \gls{mhd} simulation, and will be explored in the future.

\ack
Chang Liu would like to thank Dylan Brennan, Amitava Bhattacharjee, Eric Hollmann, Vinodh Bandaru, Mathias Hoelzl and Eric Nardon for fruitful discussion. This work was supported by the Simulation Center of electrons (SCREAM) SciDAC center by Office of Fusion Energy Science and Office of Advanced Scientific Computing of U. S. Department of Energy, under contract No. DE-SC0016268 and DE-AC02-09CH11466. The experimental research on DIII-D was supported by U.S. Department of Energy under Contract No. DE-FC02-04ER54698. This research used the high-performance computing cluster at Princeton Plasma Physics Laboratory, Traverse and Eddy clusters at Princeton University, and AiMOS cluster at the Center of Computational Innovation at Rensselaer Polytechnic Institute. This work also used the Summit cluster of the Oak Ridge Leadership Computing Facility at the Oak Ridge National Laboratory, which is supported by the Office of Science of the U.S. Department of Energy under Contract No. DE-AC05-00OR22725.

\section*{Disclaimer}

This report was prepared as an account of work sponsored by an agency of the United States Government. Neither the United States Government nor any agency thereof, nor any of their employees, makes any warranty, express or implied, or assumes any legal liability or responsibility for the accuracy, completeness, or usefulness of any information, apparatus, product, or process disclosed, or represents that its use would not infringe privately owned rights. Reference herein to any specific commercial product, process, or service by trade name, trademark, manufacturer, or otherwise, does not necessarily constitute or imply its endorsement, recommendation, or favoring by the United States Government or any agency thereof. The views and opinions of authors expressed herein do not necessarily state or reflect those of the United States Government or any agency thereof.

\appendix

\section{Derivation of divergence-free form \gls{re} convection equation}
\label{sec:divergence-free}

In the appendix we show how to derive \cref{eq:re-convection3} from \cref{eq:re-convection2}.

\begin{eqnarray}
	\frac{\partial n_{RE}}{\partial t}=-\nabla\cdot\left[n_{RE}\left(c_{RE}\mathbf{b}+\frac{\mathbf{E}\times\mathbf{B}}{B^2}\right)\right],
\end{eqnarray}
\begin{eqnarray}
	\nabla\cdot\left[n_{RE}c_{RE}\mathbf{b}\right]&=\nabla\cdot\left[n_{RE}c_{RE}\frac{\mathbf{B}}{B}\right]\\
	&=c_{RE}\mathbf{B}\cdot\nabla\left[\frac{n_{RE}}{B}\right]+\frac{n_{RE}c_{RE}}{B}\nabla\cdot\mathbf{B}\\
	&=c_{RE}\mathbf{B}\cdot\nabla\left[\frac{n_{RE}}{B}\right],
\end{eqnarray}
\begin{eqnarray}
	\nabla\cdot\left[n_{RE}\frac{\mathbf{E}\times\mathbf{B}}{B^2}\right]&=\frac{\mathbf{E}\times\mathbf{B}}{B}\cdot\nabla\left[\frac{n_{RE}}{B}\right]+\frac{n_{RE}}{B}\nabla\cdot\left[\frac{\mathbf{E}\times\mathbf{B}}{B}\right],
\end{eqnarray}
\begin{eqnarray}
	\nabla\cdot\left[\frac{\mathbf{E}\times\mathbf{B}}{B}\right]&=\mathbf{b}\cdot\left(\nabla\times\mathbf{E}\right)-\mathbf{E}\cdot\left(\nabla\times\mathbf{b}\right)\\
	&=-\frac{\partial B}{\partial t}-\mathbf{E}\cdot\left(\nabla\times\mathbf{b}\right).
\end{eqnarray}
Thus
\begin{eqnarray}
	\frac{\partial}{\partial t}\left(\frac{n_{RE}}{B}\right)&=\frac{1}{B}\frac{\partial n_{RE}}{\partial t}-\frac{n_{RE}}{B^2}\frac{\partial B}{\partial t}\\
	&=-\left(c_{RE}\mathbf{b}+\frac{\mathbf{E}\times\mathbf{B}}{B^2}\right)\cdot\nabla\left(\frac{n_{RE}}{B}\right)+\frac{n_{RE}}{B^2}\left[\mathbf{E}\cdot\left(\nabla\times\mathbf{b}\right)\right].
\end{eqnarray}

The last term can be written as
\begin{eqnarray}
	\mathbf{E}\cdot\left(\nabla\times\mathbf{b}\right)&=\mathbf{E}\cdot\left(\nabla\times\frac{\mathbf{B}}{B}\right)\\
	&=\frac{\mathbf{E}\cdot\mathbf{J}}{B}+\mathbf{E}\cdot\left(\mathbf{b}\times\frac{\nabla B}{B}\right) \label{eq:last-term}
\end{eqnarray}
Given that $\mathbf{E}=\mathbf{v}\times\mathbf{B}+\eta\left(\mathbf{J}-\mathbf{J}_{RE}\right)$, the first term in Eq.~(\ref{eq:last-term}) represents the dissipation of $|B|$ due to resistivity of Ohmic current. Since $\nabla B/B\approx -1/R \hat{\mathbf{R}}$, the second term is proportional to horizontal displacement of plasma, which is also related to the resistive decay of current.
\section*{References}
	
\bibliography{paper}

\end{document}